\documentclass[english,showpacs,preprintnumbers,superscriptaddress,amsmath,floatfix,prd]{revtex4}
\usepackage[T1]{fontenc}
\usepackage[latin1]{inputenc}
\usepackage{graphicx}
\usepackage{amssymb}

\makeatletter

\usepackage{bm}
\usepackage{psfrag}

\usepackage{babel}
\makeatother
\usepackage[dvips]{color}

\usepackage[normalem]{ulem}  % \sout{old text} for strikeout

\newcommand\g{\gamma}
\renewcommand\d{\delta}
\newcommand\e{\epsilon}

\newcommand\m{\mu}

\newcommand{\non}{\nonumber\\}

\newcommand{\be}{\begin{equation}}
\newcommand{\ee}{\end{equation}}
\newcommand{\bea}{\begin{eqnarray}}
\newcommand{\eea}{\end{eqnarray}}
\newcommand{\ba}[1]{\begin{array}{#1}}
\newcommand{\ea}{\end{array}}

\begin{document}

\title{Relativistic BCS-BEC crossover in a boson-fermion model}

\author{Jian Deng}
\email{djdddd@mail.ustc.edu.cn}
\affiliation{Department of Modern Physics, University of Science and Technology
of China, Anhui 230026, People's Republic of China}

\author{Andreas Schmitt}
\email{aschmitt@wuphys.wustl.edu}
\affiliation{Department of Physics, Washington University St Louis, MO 63130, USA}

\author{Qun Wang}
\email{qunwang@ustc.edu.cn}
\affiliation{Department of Modern Physics, University of Science and Technology
of China, Anhui 230026, People's Republic of China}

\begin{abstract}
We investigate the crossover from Bardeen-Cooper-Schrieffer (BCS) pairing
to a Bose-Einstein condensate (BEC) in a relativistic superfluid within a boson-fermion
model. The model includes, besides the fermions, separate bosonic degrees of freedom, accounting for
the bosonic nature of the Cooper pairs.
The crossover is realized by tuning the difference between the boson mass
and boson chemical potential as a free parameter.
The model yields populations of condensed and uncondensed bosons as well as gapped
and ungapped fermions throughout the crossover region for arbitrary temperatures.
Moreover, we observe the appearance of antiparticles for sufficiently large values of the
crossover parameter. As an application, we study pairing of fermions with
imbalanced populations. The model can potentially be applied to color superconductivity
in dense quark matter at strong couplings.
\end{abstract}


%\date{April 26, 2007}

\date{May 5, 2007}

\maketitle

\section{Introduction}

An arbitrarily weak attractive interaction between fermions in a many-fermion
system leads to the formation of Cooper pairs.
This phenomenon is well described within Bardeen-Cooper-Schrieffer (BCS)
theory \cite{bcs}. In this situation, Cooper pairs
are typically of a size much larger than the mean interparticle distance.
The picture changes for sufficiently
large interaction strengths. In this case, Cooper pairs become bound states, and
superfluidity is realized by a Bose-Einstein condensation (BEC)
of molecular bosons composed of two fermions. A crossover between the weak-coupling
BCS regime and the strong-coupling BEC regime is expected \cite{Eagles:1969}.

Experimentally, this crossover has been studied in systems of cold fermionic atoms in a magnetic trap,
where the coupling strength can be tuned around a Feshbach resonance with the help of an external
magnetic field \cite{coldatoms}. Recently, these studies have been extended to the case two
fermion species with imbalanced populations \cite{KetterleImbalancedSpin}. In this case, the crossover
is most likely replaced by one or more phase transitions, and the appearance of
exotic superfluids seems to be a very interesting possibility \cite{son,Gubankova:2006gj}.

Besides the nonrelativistic atomic systems, there is also a strong motivation to study the relativistic
BCS-BEC crossover. One possible realization is pion condensation, which, for large isospin densities,
crosses over into Cooper pairing of quarks and antiquarks \cite{Son:2000xc}. Another possibility
is dense quark matter which may be present in compact stars
\cite{Collins:1974ky}. Under
astrophysical conditions of densities of a few times the nuclear ground state
density and comparably small temperatures of 1 MeV and lower, quark matter is a color superconductor
\cite{Barrois:1977xd,reviews}. Analogous to electrons in a metal or alloy or fermionic atoms
in a magnetic trap, quarks form Cooper pairs due to an attractive interaction, here mediated by
gluon exchange. Because of asymptotic freedom, color superconductivity at asymptotically
large densities can be studied in a weak-coupling approach using perturbative methods within
QCD \cite{Son:1998uk,Wang:2001aq}. However, for moderate densities as present in compact stars,
the validity of these results is questionable. More phenomenological models such as the
Nambu-Jona-Lasinio (NJL) model, have
therefore been employed, mimicking the gluon exchange by a pointlike interaction between the quarks
(see Ref.\ \cite{Buballa:2003qv} and references therein).
Both QCD and NJL approaches usually are applied
within a BCS-like picture.
However, quark matter in compact stars may well be in a strong-coupling regime
where a BEC-like picture is more appropriate
\cite{Nawa:2005sb,Kitazawa:2006zp,Nishida:2005ds,Abuki:2006dv,He:2007kd}.

%The model employed in this paper is too simple to describe a realistic scenario of
%color-superconducting quark matter since it contains only a single
%fermion species and therefore is unable to account for different pairing patterns of quarks.
%Therefore, it also cannot account for a possible mismatch
%in Fermi momenta of the constituents of a Cooper pair. This is an unavoidable complication
%in realistic color and electrically neutral color superconductors \cite{Rajagopal:2005dg}.
%However, our model can certainly be extended to include these effects.

In order to describe the crossover from BCS to BEC we shall not consider a purely fermionic
model which may describe this crossover as a function of the fermionic coupling strength. We rather
set up a theory with bosonic and fermionic degrees of freedom. Here, fermions and bosons
are coupled through a Yukawa interaction and required to be in chemical equilibrium, $2\mu=\mu_b$,
where $\mu$ and $\mu_b$ are the fermion and boson chemical potentials, respectively.
We treat the (renormalized) boson mass $m_{b,r}$ and the
boson-fermion coupling $g$ as free parameters. Then,
tuning the parameter $x=-(m_{b,r}^2-\mu_b^2)/(4g^2)$
drives the system from the BCS to the BEC regime. The fermionic chemical potential shall
be self-consistently determined from the gap equation and charge conservation. This
picture is inspired by the boson-fermion model of superconductivity considered in
Ref.\ \cite{Friedberg:1989gj}, which has been used in the context of cold fermionic atoms \cite{timmermans}.
It also has possible applications for high-temperature superconductivity
\cite{domanski}. For simplicity, we shall restrict ourselves to the evaluation of
the model in a mean-field approximation.

The paper is organized as follows. In Sec.\ \ref{bf-model} we set up the simplest
version of the model, taking into account a single fermion species. We define the
crossover parameter in Sec.\ \ref{crosspara} and derive the density and gap
equations in Sec.\ \ref{densgap}. The solutions of these equations are presented in Sec.\ \ref{resultdisc}.
We consider a vanishing temperature in Sec.\ \ref{zeroT}, present the crossover at the
critical temperature in Sec.\ \ref{Tcrit}, and show results for a fixed crossover parameter and
arbitrary temperature in Sec.\ \ref{allT}. In Sec.\ \ref{unitary} we present the results
for the ratios $\Delta/\mu$ and $T_c/\Delta$. Finally, we extend our model to two fermion
species in Sec.\ \ref{imbalanced}. This extension allows us to consider pairing of
fermions with imbalanced populations, which is an unavoidable complication in quark matter
at moderate densities \cite{Rajagopal:2005dg}.

Our convention for the metric tensor is $g^{\mu\nu}=\textrm{diag}(1,-1,-1,-1)$.
Our units are $\hbar=c=k_{B}=1$. Four-vectors are denoted by capital
letters, $K\equiv K^{\mu}=(k_{0},\mathbf{k})$ with $k=|\mathbf{k}|$. Fermionic Matsubara frequencies are
$\omega_n=ik_0=(2n+1)\pi T$, while bosonic ones are $\omega_n=ik_0=2n\pi T$ with the temperature $T$ and
$n$ an integer.

\section{The boson-fermion model for a relativistic superfluid}
\label{bf-model}

\subsection{Setting up the model}

We use a model of fermions and composite bosons coupled to each other by a Yukawa interaction.
The Lagrangian is given by a free fermion part ${\cal L}_f$, a free boson part ${\cal L}_b$
and an interaction part ${\cal L}_I$,
\begin{equation}
\mathcal{L}=\mathcal{L}_{f}+\mathcal{L}_{b}+\mathcal{L}_{I} \, ,\label{eq:lag}
\end{equation}
with
\begin{subequations}
\begin{eqnarray}
\label{free-term}
\mathcal{L}_{f} & = & \overline{\psi}(i\g^\mu\partial_\mu+\g_0\mu-m)\psi \, ,  \\
\mathcal{L}_{b} & = & |(\partial_t-i\mu_b)\varphi|^2-|\nabla\varphi|^2-m_b|\varphi|^2 \, , \\
\mathcal{L}_{I} & = & g(\varphi\overline{\psi}_{C}i\gamma_{5}\psi+\varphi^*
\overline{\psi}i\gamma_{5}\psi_{C}) \, . \label{eq:lagrangian-fluc1}
\end{eqnarray}
\end{subequations}
The fermions are described by the spinor $\psi$, while the bosons are given by the complex scalar field
$\varphi$. The charge conjugate spinors are defined by $\psi_{C}=C\overline{\psi}^{T}$
and $\overline{\psi}_{C}=\psi^{T}C$ with $C=i\gamma^{2}\gamma^{0}$. The fermion (boson) mass is denoted by
$m$ ($m_b$). We choose the boson chemical potential to be twice the fermion chemical potential,
\be \label{mub}
\mu_b = 2\mu \, .
\ee
Therefore, the system is in chemical equilibrium with respect to the
conversion of two fermions into one boson and vice versa. This allows us to model the transition
from weakly-coupled Cooper pairs made of two fermions into
a molecular difermionic bound state, described as a boson. The interaction term accounts for a
local interaction
between fermions and bosons with coupling constant $g$. In order to describe BEC of the bosons, we
have to separate the zero mode of the field $\varphi$ \cite{kapusta}. Moreover, we shall
replace this zero-mode by its expectation value
\be
\phi\equiv \langle\varphi_0\rangle
\ee
and neglect the interaction between the fermions and the {\em non-zero} boson modes.
This corresponds to the mean-field approximation.
Then, with the Nambu-Gorkov spinors
\be
\Psi=\left(\begin{array}{c}
\psi\\
\psi_{C}\end{array}\right)\, ,\qquad \overline{\Psi}=(\overline{\psi},\overline{\psi}_{C}) \, ,
\ee
the Lagrangian can be written as
\be
{\cal L} = \frac{1}{2}
\overline{\Psi} {\cal S}^{-1} \Psi +
%\mathcal{L}_{b}
[\mu ^2_b-m_b^2]|\phi|^2 + |(\partial_t-i\mu_b)\varphi|^2
-|\nabla\varphi|^2 - m_b^2|\varphi|^2
\, .
\label{lagrange}
\ee
Note that we have dropped the mixing terms of zero and non-zero boson modes
since they vanish when carrying out the path intergal.
Here ${\cal S}^{-1}$ is the inverse fermion propagator which reads
in momentum space
\begin{eqnarray}
{\cal S}^{-1}(P) & = & \left(\begin{array}{cc}
P_{\mu}\gamma^{\mu}+\mu\gamma_{0}-m & 2ig\gamma_{5}\phi^*\\
2ig\gamma_{5}\phi & P_{\mu}\gamma^{\mu}-\mu\gamma_{0}-m\end{array}\right) \, .
\label{eq:s-inverse}
\end{eqnarray}
It is instructive to compare this form of the propagator to the corresponding one in a purely
fermionic model, see for instance Ref.\ \cite{Kitazawa:2006zp}. As expected, the Bose condensate
is related to the diquark condensate $\Delta$,
\be \label{Delta}
\Delta = 2g\phi \, .
\ee
As we shall see below, cf.\ Eq.\ (\ref{fermions}), $\Delta$ is the energy gap in the quasi-fermion
excitation spectrum.
In a purely fermionic model, $\Delta = 2 G \langle \overline{\psi}_Ci\g_5\psi\rangle$, where $G$ is
the coupling constant related to the interaction between the fermions. Note that $G$ has
mass dimension $-2$, while our boson-fermion coupling $g$ is dimensionless. Therefore, $g$
does not play the role of the crossover parameter, as $G$ does in the fermionic model. We shall
explain this in more detail in Sec.\ \ref{crosspara}.

\subsection{Thermodynamic potential}

In order to obtain the thermodynamical potential density $\Omega$, we compute the
partition function
\begin{eqnarray}
{\cal Z} & = & \int[d\Psi][d\overline{\Psi}][d\varphi][d\varphi^*]
\exp\left[\int_{0}^{1/T}d\tau d^{3}x\,\mathcal{L}\right] \, ,
\end{eqnarray}
where $T$ is the temperature, and ${\cal L}$
is the Lagrangian in the mean field approximation given in Eq.\ (\ref{lagrange}).
The thermodynamic potential density is then obtained from $\Omega = -T/V\,\ln{\cal Z}$,
where $V$ is the volume of the system.  One obtains
after performing the path integral and the sum over Matsubara frequencies,
\begin{eqnarray}
\Omega & = & -\sum_{e=\pm}\int\frac{d^{3}k}{(2\pi)^{3}}\left\{ \e_k^e +
2T\ln\left[1+\exp\left(-\frac{\e_k^e}{T}\right)\right]\right\} \non
&& +\; \frac{(m_b^{2}-\mu _b^2)\Delta^2}{4g^2} +
\frac{1}{2}\sum_{e=\pm}\int\frac{d^{3}k}{(2\pi)^{3}}\left\{
\omega_k^e+2T\ln\left[1-\exp\left(-\frac{\omega_k^e}{T}
\right)\right]\right\} \, .
\label{eq:tot-the-pot}
\end{eqnarray}
We have used Eq.\ (\ref{Delta}) and denoted the quasi-particle energy
for fermions ($e=+1$) and antifermions ($e=-1$) by
\be \label{fermions}
\e_k^e = \sqrt{(\e_{k0}-e\mu)^2+\Delta^2} \, , \qquad \e_{k0} = \sqrt{k^2+m^2} \, ,
\ee
and the (anti)boson energy by
\be \label{omegak}
\omega_k^e = \sqrt{k^2+m_b^2} - e\mu_b \, .
\ee
Furthermore, we have assumed $\Delta$ (and thus $\phi$) to be real.

\subsection{Crossover parameter}
\label{crosspara}

We shall now define the crossover parameter whose variation carries the system from the BCS to the BEC
regime. Let us first recall the corresponding crossover parameter in a
purely fermionic model. In this case, the fermionic coupling $G$ has to be renormalized. This is in contrast
to the weak-coupling regime where the gap equation is well-defined with the bare coupling $G$ (a natural
cutoff is provided by the Debye frequency in the non-relativistic case; in QCD, the gap is a function
of momentum and peaks around the Fermi surface, providing a regular behavior of the gap equation).
The renormalized coupling is proportional to the scattering length.
In the context of cold fermionic
atoms, the scattering length is the physical quantity which can be controlled upon tuning the
external magnetic
field. For the relativistic case, see Ref.\ \cite{Abuki:2006dv} for the relation between the
renormalized coupling and the scattering length. The definition of the crossover
parameter in the present model goes along the same lines. Instead of a renormalized coupling
we introduce the renormalized boson mass
\be \label{mr}
m_{b,r}^2 = 4g^2\left.\frac{\partial\Omega}{\partial \Delta^2}\right|_{\Delta=\mu=T=0}
=m_b^2 -4g^2\int\frac{d^3k}{(2\pi)^3}\frac{1}{\e_{k0}} \, .
\ee
This allows us to define the (renormalized) crossover parameter
\be \label{defx}
x\equiv -\frac{m_{b,r}^2-\mu _b^2}{4g^2} \, .
\ee
The parameter $x$ can be varied from negative values with large modulus (BCS) to large positive values
(BEC). In between, $x=0$ is the unitary limit \cite{chang,Carlson:2005kg,Nishida:2006br}.
Therefore, $1/x$ behaves similar to the scattering length:
the BCS (BEC) limit is approached via $1/x\uparrow 0$ ($1/x\downarrow 0$) while the unitary
regime corresponds to $1/x=\pm\infty$.

We may thus write the thermodynamical potential in terms of the parameters $(x,g)$ instead of the
original pair $(m_b,g)$. To this end, we have to express the bare boson mass $m_b$
in terms of $x$ and $g$. With the help of Eqs.\ (\ref{mr}) and (\ref{defx})
we find
\be \label{baremass}
m_b^2-\mu _b^2 = 4g^2\left(\int\frac{d^3k}{(2\pi)^3}\frac{1}{\e_{k0}}- x\right) \equiv
4g^2(x_0-x) \, .
\ee
For sufficiently small fermion masses, $m\ll \Lambda$, where $\Lambda$ is the cutoff in the momentum
integrals, we have
\be
x_0 \simeq \frac{\Lambda^2}{4\pi^2} \, .
\ee
One sees from Eq.\ (\ref{baremass}) that $x_0$ is an upper limit for $x$ in order to
ensure non-negative bosonic occupation numbers. Moreover, in the limit of large $x\to x_0$
the boson chemical potential approaches the (bare) boson mass, $\mu_b \to m_b$. The condition
of Bose-Einstein condensation in a free bosonic system with fixed bosonic charge is $\mu_b=m_b$.
In the present model, however, we shall observe a nonzero Bose condensate also for $\mu_b<m_b$,
corresponding to $x<x_0$.

Having defined the crossover parameter $x$ and its definition range $x\in [-\infty,x_0]$,
we note that, within our simple model, we are left with
the second free parameter $g$. We shall discuss below how the choice of $g$ effects the behavior
of the system in the BCS-BEC crossover. For most of our results we shall, however, use a single fixed
value of $g$.

\subsection{Densities and gap equation}
\label{densgap}

Next, we derive the charge conservation equation and the gap equation which shall later be solved
numerically. The total charge density
\be \label{defdensity}
n = -\frac{\partial\Omega}{\partial \mu}
\ee
can, using Eq.\ (\ref{eq:tot-the-pot}), be written as
\be \label{density}
n = n_F + n_0 + n_B  \, .
\ee
Here, the fermionic contribution is given by
\be \label{fermiantifermi}
n_F \equiv 2\sum_{e=\pm}e\,\int\frac{d^{3}k}{(2\pi)^{3}}\frac{\xi_k^e}{2\e_k^e}
\left[f_F(\e_k^e) - f_F(-\e_k^e)\right] \, ,
\ee
where we abbreviated
\be
\xi_k^e\equiv \e_{k0}-e\mu \, ,
\ee
and $f_F$ is the Fermi distribution function, $f_F(x) = 1/[\exp(x/T)+1]$. The factor 2 in front of the
sum in Eq.\ (\ref{fermiantifermi}) originates from the two spin degrees of freedom. From
Eq.\ (\ref{fermiantifermi}) one recovers the limit case of a free Fermi gas at zero temperature,
\be
n_F(\Delta=T=0)=\frac{(\mu^2-m^2)^{3/2}}{6\pi^2}[\Theta(\mu-m)-\Theta(-\mu-m)] \, .
\ee
The condensate density is
\be
\label{condensate}
n_0 \equiv \frac{\mu _b\,\Delta^2}{g^2} \, ,
\ee
and the thermal boson contribution is
\be \label{boseantibose}
n_B\equiv 2\sum_{e=\pm}e\,\int\frac{d^{3}k}{(2\pi)^{3}} f_B(\omega_k^e) \, ,
\ee
where $f_B$ is the Bose distribution function, $f_B(x)=1/[\exp(x/T)-1]$.
The factor 2 in the boson densities originates from Eq.\ (\ref{mub}), i.e., from the fact
that each boson is composed of two fermions. For the following, let us also define
the charge fractions
\be
\rho_{B/F}\equiv \frac{n_{B/F}}{n} \, , \qquad \rho_0\equiv \frac{n_0}{n} \, ,
\ee
and an effective Fermi momentum $p_F$ through
\be
\label{eff-pf}
n=\frac{p_F^3}{3\pi^2} \, .
\ee
The effective Fermi energy is then given by $\epsilon _F\equiv \sqrt{p_F^2+m^2}$.
The various densities appearing on the right-hand side
of Eq.\ (\ref{density}) are interpreted as follows. The fermions
that contribute to $n_F$ are,
for temperatures below the superfluid transition temperature,
constituents of weakly coupled Cooper pairs.
For temperatures larger than the transition temperature, $n_F$
corresponds to free fermions.
The bosons that contribute to the boson density $n_B$ are, for all temperatures, {\em uncondensed} molecular
bound states, composed of two fermions. Condensation of these pairs can only occur below
the transition temperature and
results in a nonzero condensate density $n_0$.

In order to find the gap equation in the case of a fixed charge density, we have to minimize the
free energy density
\be
F = \Omega + \mu n \, .
\ee
Here, $\mu$ is an implicit function of $n$ (and of the gap $\Delta$) through Eq.\ (\ref{defdensity}).
Minimization with respect to $\Delta$ yields
\be \label{minimize}
0 = \frac{dF}{d\Delta} = \frac{\partial\Omega}{\partial \Delta} + \frac{\partial\Omega}{\partial \mu}
\frac{\partial\mu}{\partial \Delta} + n\,\frac{\partial\mu}{\partial \Delta} =
\frac{\partial\Omega}{\partial \Delta} \, ,
\ee
where Eq.\ (\ref{defdensity}) has been used. For $\Delta\neq 0$ Eq.\ (\ref{minimize}) reads
\be\label{gapequation}
-x = \sum_{e=\pm}\int\frac{d^{3}k}{(2\pi)^{3}}\left(\frac{1}{2\e_k^e}\tanh\frac{\e_k^e}{2T}
-\frac{1}{2\e_{k0}}\right) \, .
\ee
Note that the density $n$ in Eq.\ (\ref{density})
was obtained by taking the derivative with respect to
$\mu$ at fixed $m_b$ (not at fixed $x$). This is necessary to obtain a nonzero condensate
contribution $n_0$. Also the equivalence of $dF/d\Delta$ and $\partial\Omega/\partial\Delta$ in
Eq.\ (\ref{minimize}) is obtained under this premise. At fixed $m_b$ we get $\mu$ by solving
Eqs.\ (\ref{density}) and (\ref{gapequation}). We can then obtain $x$ from $m_b$ and $\mu$
via Eq.\ (\ref{defx}). In this way we have a one-to-one mapping between $m_b$ and $x$.
Here we should emphasize that the current case is different from that $x$ is fixed from
the very beginning before Eq.\ (\ref{density}) is derived.

%We have checked numerically that the results are virtually indistinguishable
%from considering a fixed $x$ (which renders the minimization of
%the free energy more complicated).

\section{Results and discussion}
\label{resultdisc}

The two coupled equations (\ref{density}) and (\ref{gapequation})
with the definitions (\ref{fermiantifermi}), (\ref{condensate}), and (\ref{boseantibose})
shall be used in the following to determine the gap $\Delta$,
and the chemical potential $\mu$ as
functions of the crossover parameter $x$, see Eq.\ (\ref{defx}), and the temperature $T$ at fixed
effective Fermi momentum $p_F$, fermion mass
$m$, and boson-fermion coupling $g$. The solution [$\Delta(x,T)$, $\mu(x,T)$] can then, in turn, be used to
compute the densities of fermions and bosons in the $x$-$T$ plane. We shall
present results for the zero-temperature case, Sec.\ \ref{zeroT}, and at the critical temperature $T_c$,
Sec.\ \ref{Tcrit}. Then, we show results for a fixed $x$
and arbitrary temperature $T$, Sec.\ \ref{allT}.
Throughout these subsections, we shall fix
\be \label{parameters}
\frac{p_F}{\Lambda} = 0.3\, , \qquad \frac{m}{\Lambda} = 0.2 \, , \qquad g=4 \, .
\ee
In Sec.\ \ref{unitary} we present the ratios $\Delta_0/\mu_0$ and $T_c/\Delta_0$ for different values
of $m$ and $g$ as a function of $x$. Here and in the following,
we use the subscript 0 at $\Delta$ and $\mu$ to denote the zero-temperature value.

\subsection{Zero temperature}
\label{zeroT}

\begin{figure*} [ht]
\begin{center}
\hbox{\includegraphics[width=0.45\textwidth]{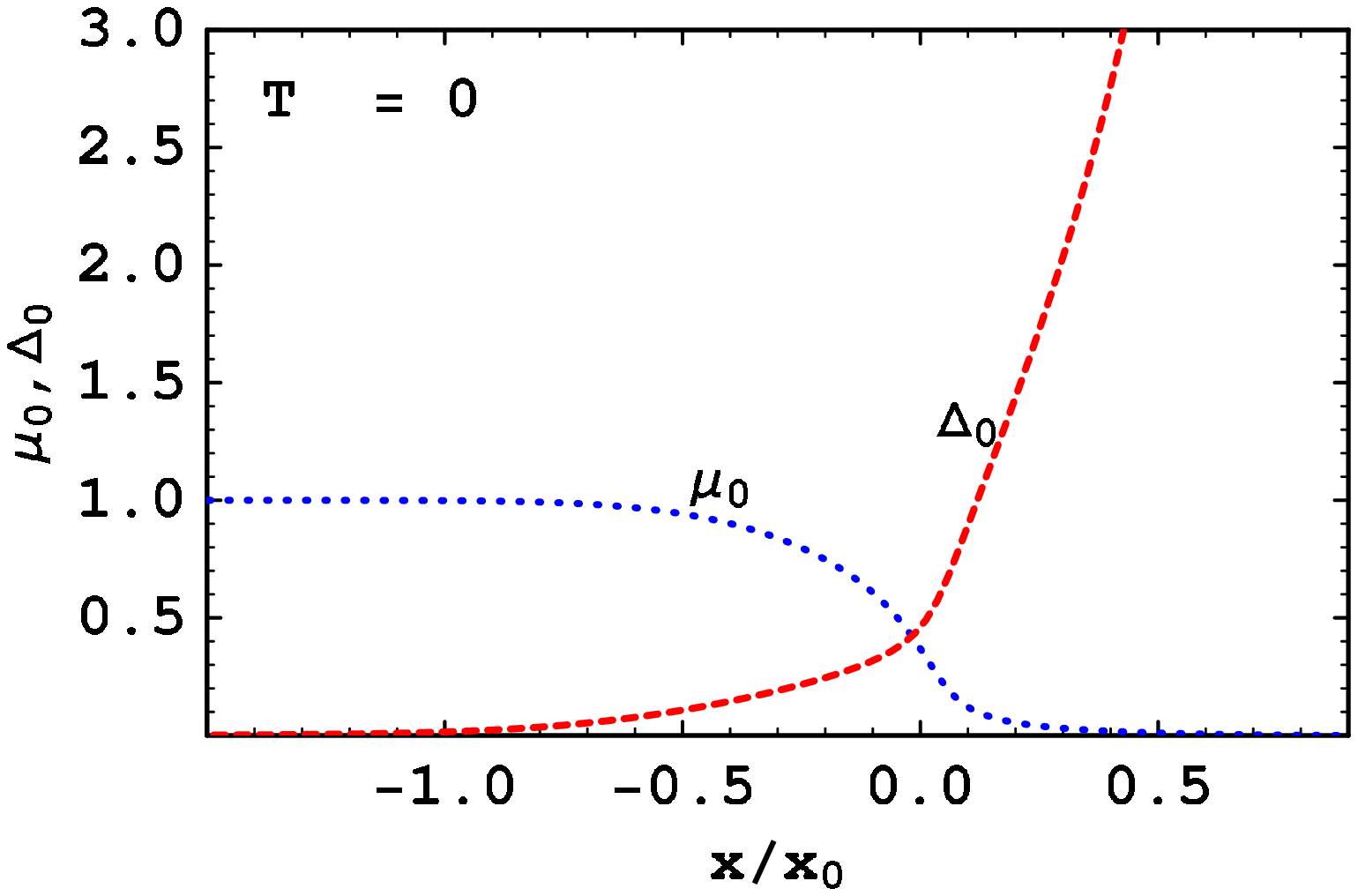}\hspace{0.5cm}
\includegraphics[width=0.46\textwidth]{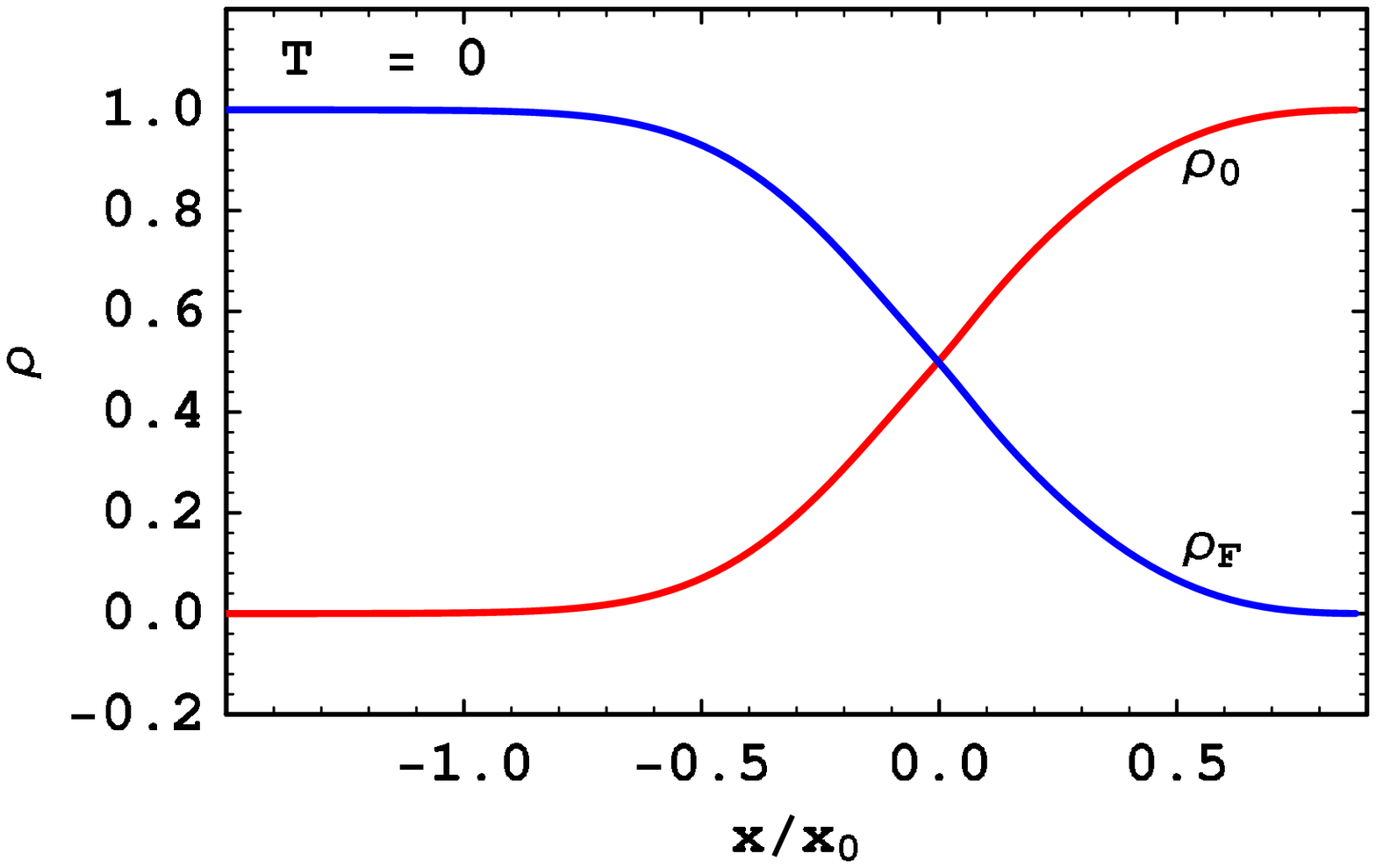}}
%\vspace{0.5cm}
\caption{(Color online)
Crossover at zero temperature from the BCS regime (small $x$) to the BEC regime (large $x$).
Left panel: fermion chemical potential $\mu _0$ (blue dotted)
and gap $\Delta _0$ (red dashed) in units of effective Fermi energy $\e _F$.
Right panel: condensate fraction (red solid), fermion fraction (blue solid).}
\label{figzeroT}
\end{center}
\end{figure*}

For $T=0$, there are no thermal bosons, $n_B=0$, and Eqs.\ (\ref{density}) and (\ref{gapequation}) become
\begin{subequations}
\bea \label{densityzero}
n &=& n_F + n_0 \, , \\ \nonumber \\
-x &=& \sum_{e=\pm}\int\frac{d^{3}k}{(2\pi)^{3}}\left(\frac{1}{2\e_k^e}
-\frac{1}{2\e_{k0}}\right) \, ,\label{gapequationzero}
\eea
\end{subequations}
with the zero-temperature expressions for the fermion densities
\be \label{fermizeroT}
n_F \equiv 2\sum_{e=\pm}e\,\int\frac{d^{3}k}{(2\pi)^{3}}\frac{\xi_k^e}{2\e_k^e} \, .
\ee
The numerical results for the solution of the coupled equations (\ref{densityzero}) and
(\ref{gapequationzero}) are shown in Fig. \ref{figzeroT}.
The left panel shows the fermion chemical
potential $\mu _0$ and the gap $\Delta _0$ as functions of $x$.
In the weak-coupling regime (small $x$)
we see that the chemical potential is given by the Fermi energy, $\mu_0 = \e_F$.
For the given parameters, $\e_F/\Lambda \simeq 0.36$. The chemical potential
decreases with increasing $x$ and approaches zero in the far BEC region.
The gap is exponentially small in the
weak-coupling region, as expected from BCS theory. It becomes of the order of the chemical potential
around the unitary limit, $x=0$, and further increases monotonically for positive $x$.
In the unitary limit, we have $\mu /\epsilon_F \simeq 0.37$, while
in nonrelativistic fermionic models $\mu /\epsilon_F \simeq 0.4-0.5$ was obtained
\cite{chang,Carlson:2005kg,Nishida:2006br}.

The corresponding fermion and boson densities are shown in the right panel of
Fig. \ref{figzeroT}.
These two curves show the crossover: at small $x$
all Cooper pairs are resonant states, which is characterized by a purely fermionic density, $n=n_F$;
at large $x$, on the other hand,
Cooper pairs are bound states and hence there are no fermions in the system. The charge density
is rather dominated by a bosonic condensate, $n=n_0$.
The crossover region is located around $x=0$. We can characterize this region quantitatively
as follows. We write the boson mass as $m_b = 2m-E_{\rm bind}$. Then, a bound state appears for
positive values of the binding energy $E_{\rm bind}$, i.e., for $2m>m_b$. With Eq.\ (\ref{baremass})
this inequality reads
\be
\mu_0^2 < m^2 -g^2(x_0-x) \, .
\ee
Since $\mu_0$ is a monotonically decreasing function of $x$, this relation suggests
(note that $x_0-x>0$ by construction): $(i)$ for sufficiently large $x<x_0$ bound states appear
for any fixed $g$ $(ii)$ the larger $g$ the ``later''
(= larger values of $x$) bosonic states appear.
We have confirmed these two statements numerically by using different values
of $g$. The value $g=4$ is chosen such that there is an approximately balanced coexistence
of fermions and bosons at $x=0$, $n_F \simeq n_0$, as can be seen in the
right panel of Fig.\ \ref{figzeroT}.

One may ask whether there is a contribution of antifermions to the total fermion charge.
In the BCS regime there is a Fermi surface given by $\mu>0$ and antifermion excitations are obviously
suppressed. However, during the crossover, $\mu$ decreases and there might be the possibility of
the appearance of antifermions. The contributions of fermions and antifermions to
the total fermion charge seem to be given by the terms
$e=+1$ and $e=-1$ in Eq.\ (\ref{fermizeroT}).
A separate discussion of these terms is not straightforward because
they contain divergent contributions which cancel in the sum but not in each term
separately. Thus, a renormalization of both terms would be required. In the BCS regime,
$x\to-\infty$, vacuum contributions $\propto\Lambda^3$ have to
be subtracted. For nonzero values of the gap, however, more divergent terms appear, involving
powers of both the cutoff and the gap. (Note that this problem is not unlike the one
encountered in Ref.\ \cite{Alford:2005qw}, where medium-dependent counter terms were introduced in the
calculation of the Meissner mass. In fact, we shall choose a similar renormalization
in the following calculation of the energy density.)

In any case, separate charges of fermions and antifermions are not measurable quantities
since any potential experiment would solely measure the total charge. Therefore, we
shall describe the onset of a nonzero antiparticle population in terms of the energy density.
In this quantity, we expect the contributions from particles and antiparticles to add up, in
contrast to the charge density where the contributions (partially) cancel each other.

\begin{figure*} [ht]
\begin{center}
\hbox{\includegraphics[width=0.33\textwidth]{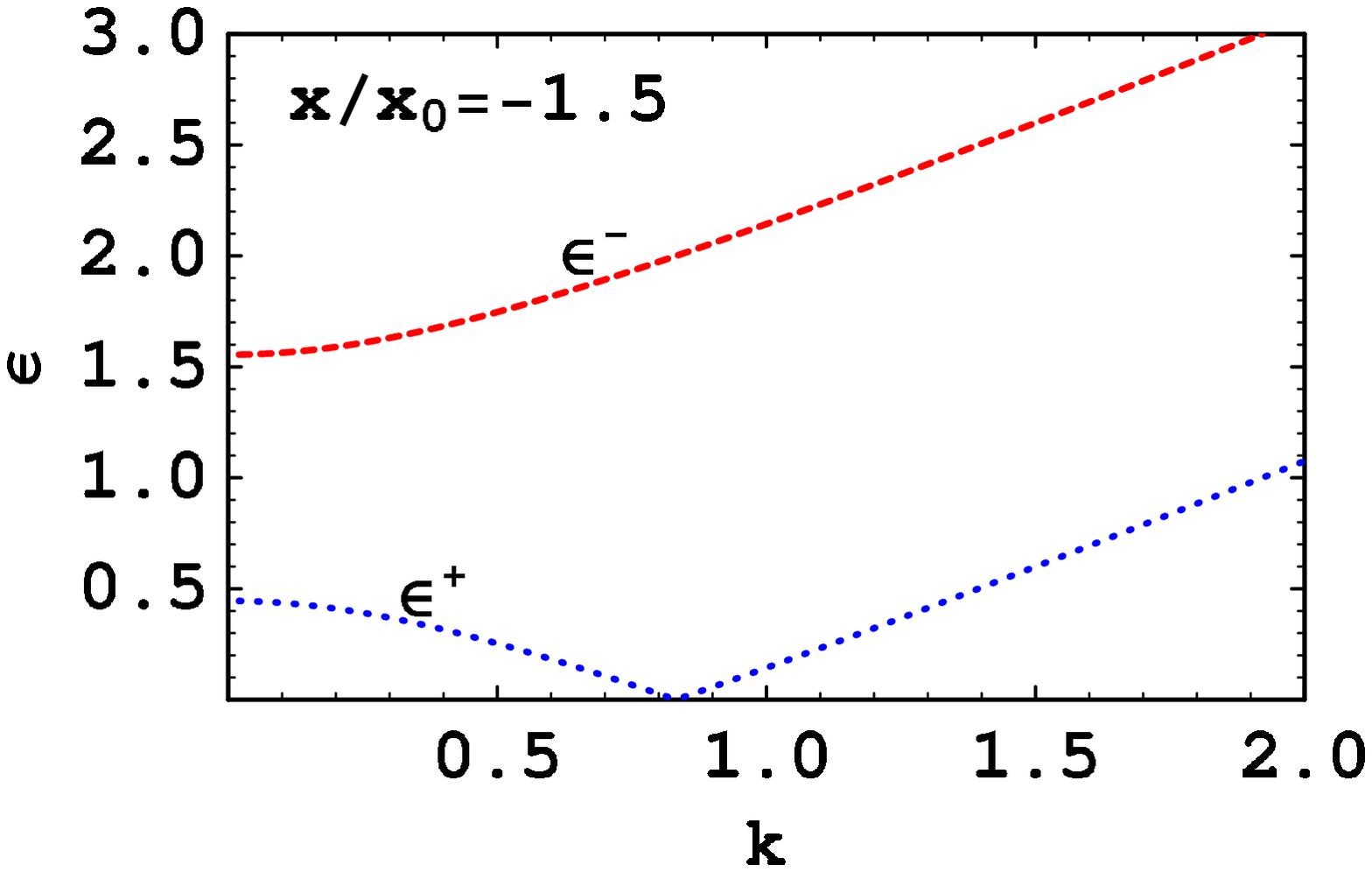}
\includegraphics[width=0.33\textwidth]{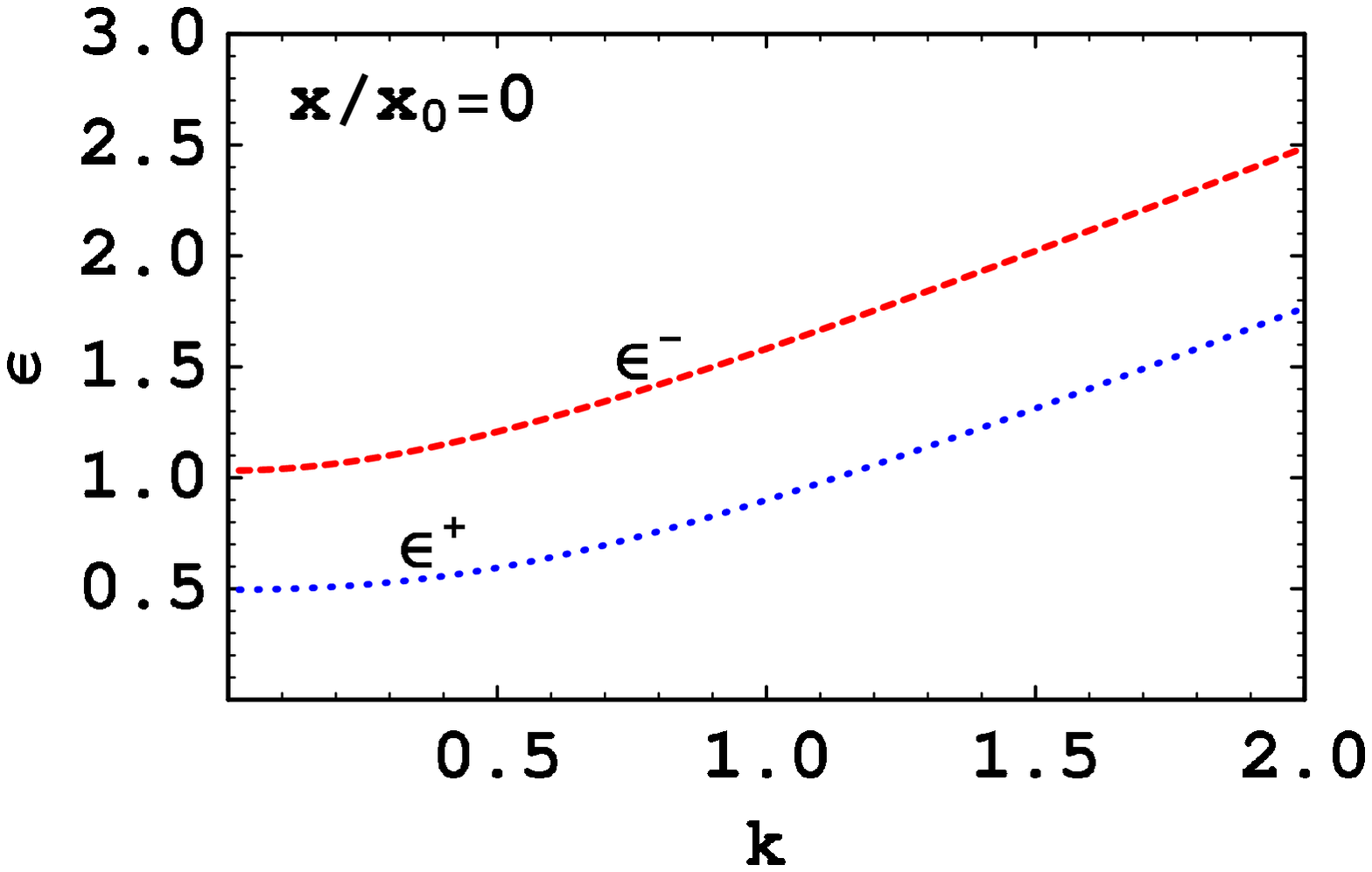}
\includegraphics[width=0.33\textwidth]{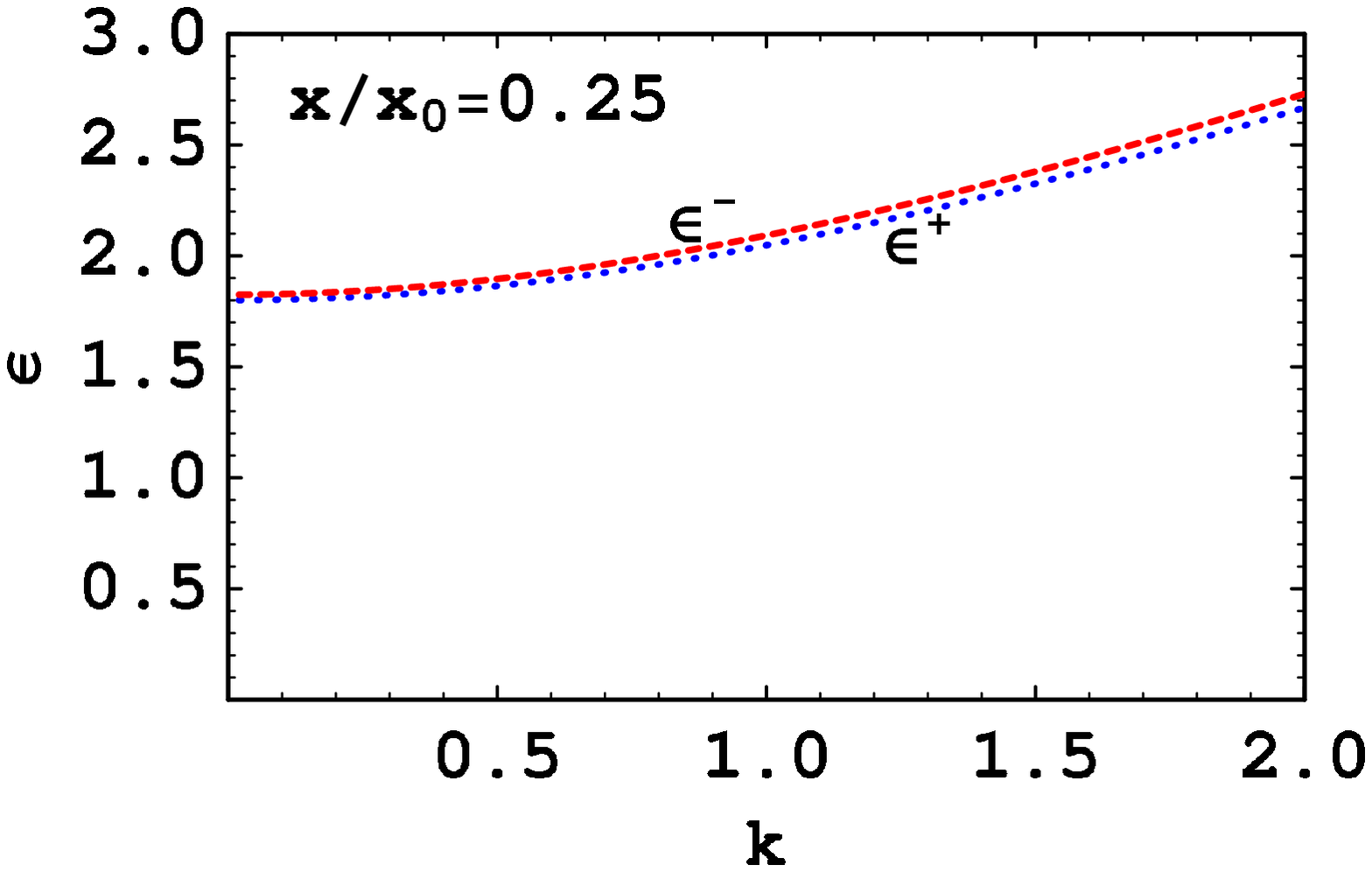}}
%\vspace{0.5cm}
\caption{(Color online)
Fermion and antifermion excitation energies $\e_k^+$ and $\e_k^-$
as defined in Eq.\ (\ref{fermions}) for three different values of the crossover parameter $x/x_0$
at $T=0$ as a function of the momentum $k$ (both $\e_k^e$ and $k$ are given in units of $\e _F$).
In the BCS regime (left panel) the energy gap is small and
the fermion excitations are well separated from antifermion excitations.
Both excitations approach each other in the unitary regime (middle panel), and become
indistinguishable in the far BEC regime (right panel). Note in particular that the minimum of the
antiparticle excitation is not a monotonic function of $x$.}
\label{figanti}
\end{center}
\end{figure*}

Let us first discuss the quasi-fermion and quasi-antifermion excitation energies given by
$\e_k^+$ and $\e_k^-$ from Eq.\ (\ref{fermions}). Inserting the numerical solutions for $\mu$
and $\Delta$ into these energies results in the curves shown in Fig.\ \ref{figanti}. These
excitation energies show that, for large values of $x$, quasi-fermions and
quasi-antifermions become degenerate due to the vanishingly small chemical potential.
Because of the large energy gap, we expect neither quasi-fermions nor quasi-antifermions to be
present in the system.

\begin{figure*} [ht]
\begin{center}
\hbox{\includegraphics[width=0.45\textwidth]{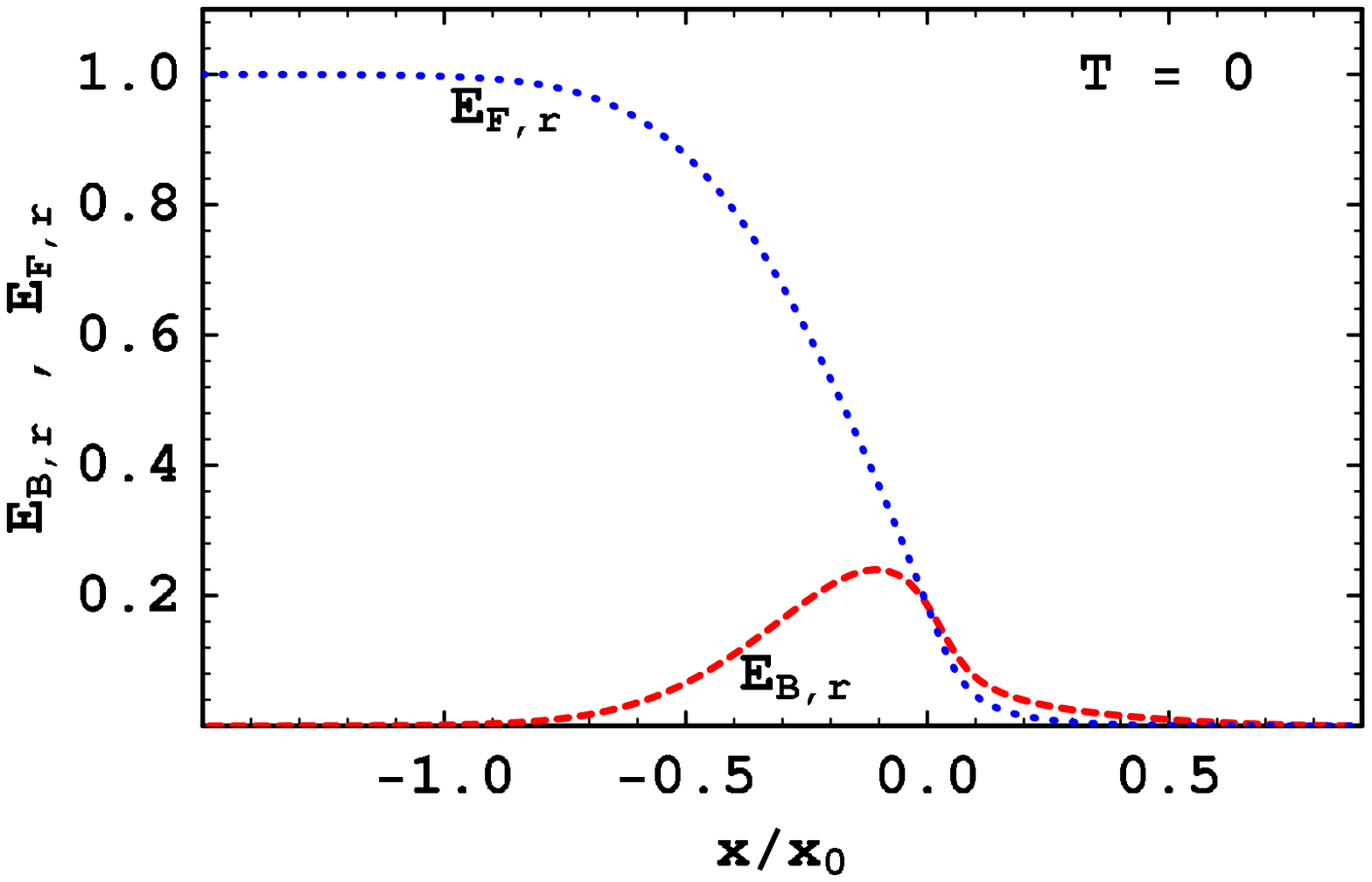}\hspace{0.5cm}
\includegraphics[width=0.45\textwidth]{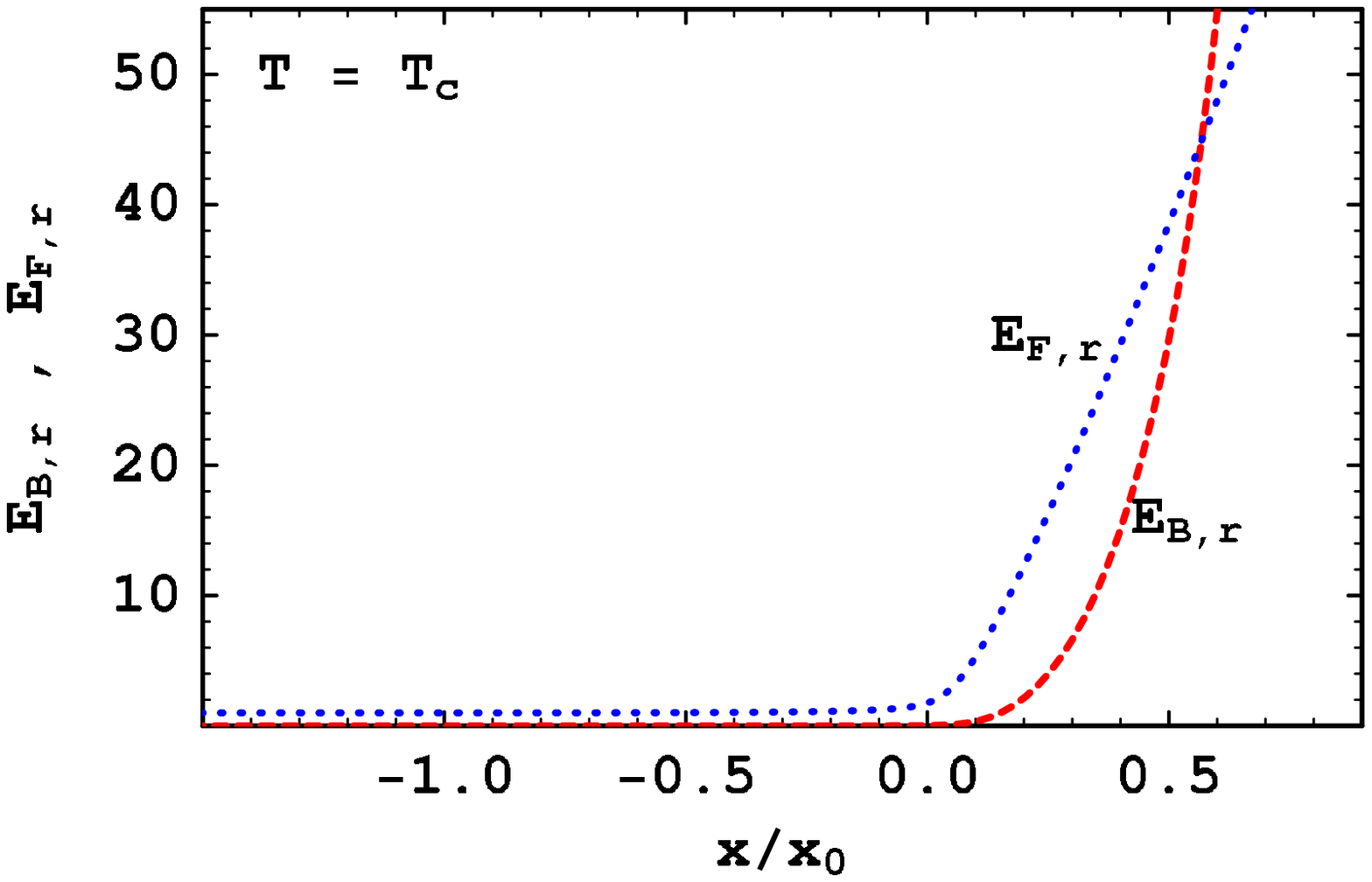}}
%\vspace{0.5cm}
\caption{(Color online)
Energy density of fermions and bosons in units of $\e_F n$ at $T=0$ (left panel) and at $T=T_c$ (right panel).
In the BCS regime, $E_F=\e_F n$ and $E_B=0$ for all temperatures.
The large fermionic and bosonic energy densities in the BEC regime at $T=T_c$
indicate the occupation of (quasi-)antiparticle modes. }
\label{figenergy}
\end{center}
\end{figure*}

This statement can be made more precise and generalized to nonzero temperatures
upon considering the energy density $E$. Using the thermodynamic
potential density $\Omega$ from Eq.\ (\ref{eq:tot-the-pot}) and the entropy density
$S=-\partial\Omega/(\partial T)$ we have $E=\Omega + \mu n + TS$. We obtain
\be
E= E_F + E_B \, ,
\ee
with the fermionic and bosonic contributions
\begin{subequations}
\bea
E_F &=& -\sum_e\int\frac{d^3k}{(2\pi)^3}\,\e_k^e[1-2f_F(\e_k^e)] + \mu\,n_F \, , \label{ef} \\
E_B &=& \frac{1}{2}\sum_e\int\frac{d^3k}{(2\pi)^3}\,\omega_k^e[1+2f_B(\omega_k^e)] +
(x_0-x)\Delta^2 + \mu(n_0+n_B) \, .
\eea
\end{subequations}
The renormalization of the fermionic part can now be chosen such that there are no
quasi-particles at $T=n_F=0$, in accordance with the above argument. Hence we subtract
the ``vacuum contribution'' $E_F(T=n_F=0)$ to obtain the renormalized energy density
\be \label{energyfermi}
E_{F,r} = 2\sum_e\int\frac{d^3k}{(2\pi)^3}\,\e_k^e f_F(\e_k^e) + \mu\,n_F  \, .
\ee
For $T=0$, only the second term survives (remember that $\e_k^e>0$), and $E_F$ behaves as
shown in the left panel of Fig.\ \ref{figenergy}. For nonzero temperatures, however,
we see that there is a nonzero fermionic energy density even for $n_F=0$. This is
related to the excitation of quasi-antifermions, as we shall discuss in the next subsection.

For the bosonic energy density, we subtract the analogous vacuum part
$E_B(T=n_0+n_B=0)$. Hence we obtain the renormalized energy density
\be \label{energybose}
E_{B,r}= \sum_e\int\frac{d^3k}{(2\pi)^3}\,\omega_k^e f_B(\omega_k^e) + \mu(n_0 +n_B)
\, .
\ee
At $T=0$, we have $E_{B,r}=\mu n_0$, which is shown in the left panel of Fig.\ \ref{figenergy}.
We see that the bosonic energy density vanishes in the BCS regime because there
is no Bose condensate in this case, $n_0=0$. In the far BEC regime, where $n_0\neq 0$,
the energy density vanishes too because the boson chemical potential, since coupled to
the fermion chemical potential, vanishes. Only in the crossover region, where both the condensate
and the chemical potential are nonzero, the energy density is nonvanishing.

\subsection{Critical temperature}
\label{Tcrit}

In this section, we calculate the critical temperature $T_{c}$ and the corresponding
particle densities as functions of $x$.
Upon setting $\Delta=0$ in the charge density equation (\ref{density}) and the gap equation
(\ref{gapequation}) one obtains
\begin{subequations}\label{gaprhoTc}
\begin{eqnarray}
n & = & n_F + n_B \, ,
\label{eq:num-03}\\ \nonumber \\
-x &=&  \sum _{e=\pm}
\int\frac{d^{3}k}{(2\pi)^{3}}\left(\frac{1}{2\xi^e_k}
\tanh\frac{\xi_k^e}{2T_{c}}-\frac{1}{2\e_{k0}}\right) \, ,\label{eq:gap-03}
\end{eqnarray}
\end{subequations}
with the fermion density
\be
n_F\equiv 2\sum_{e=\pm}e\,\int\frac{d^{3}k}{(2\pi)^{3}} f_F(\xi_k^e) \, ,
\ee
and the boson density given by Eq.\ (\ref{boseantibose}).
Strictly speaking, the original gap equation (\ref{gapequation}) is only valid for nonzero
$\Delta$ (in its derivation, one has to divide by $\Delta$). Therefore, Eq.\ (\ref{eq:gap-03})
has to be understood as a limit for approaching the critical temperature from below,
$T\uparrow T_c$, i.e., for infinitesimally small $\Delta$.
Eqs.\ (\ref{eq:num-03}) and (\ref{eq:gap-03}) can now be used to determine $T_{c}$ and the corresponding
chemical potential $\mu_{c}$.

\begin{figure*} [ht]
\begin{center}
\hbox{\includegraphics[width=0.45\textwidth]{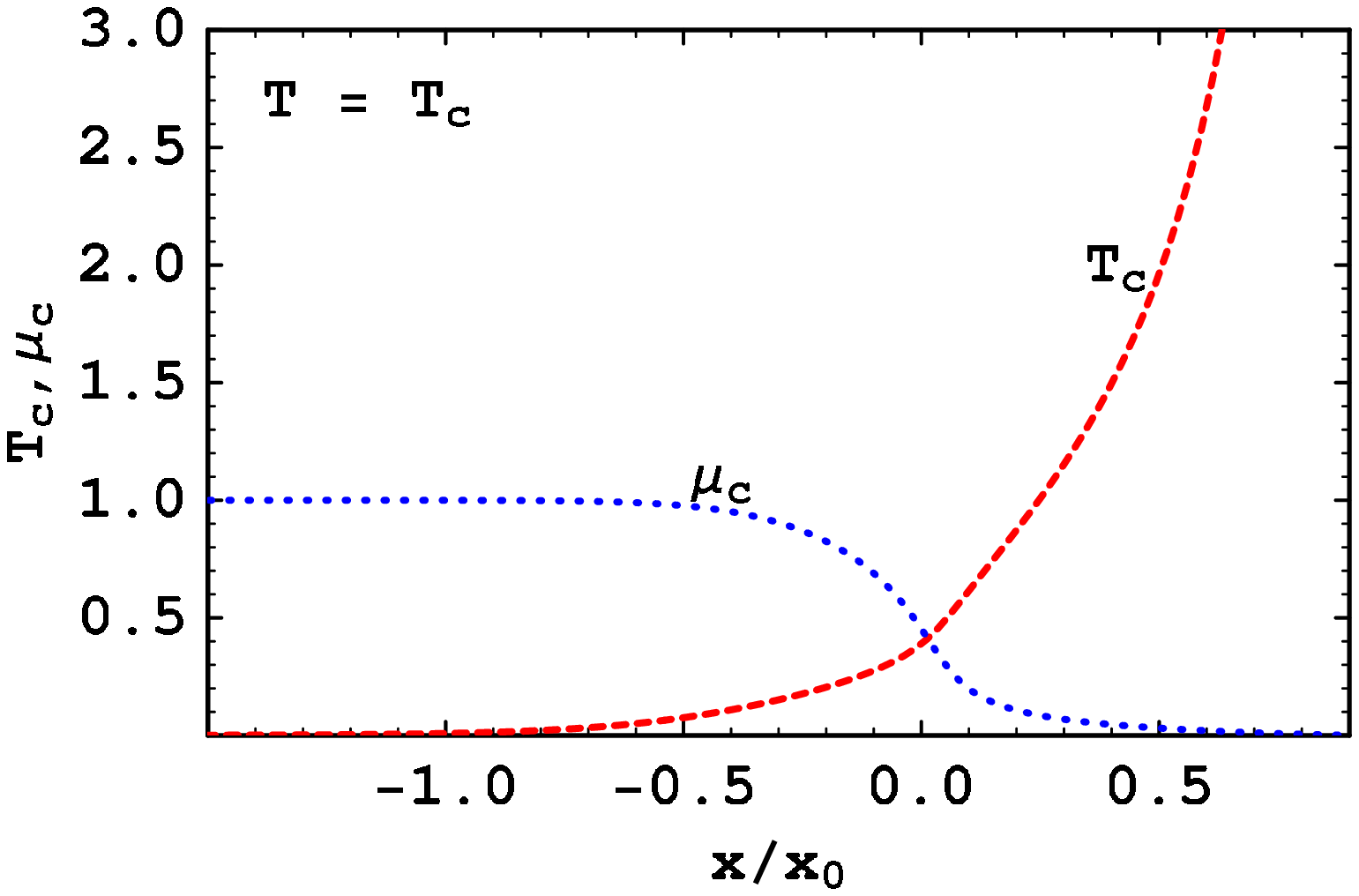}\hspace{0.5cm}
\includegraphics[width=0.45\textwidth]{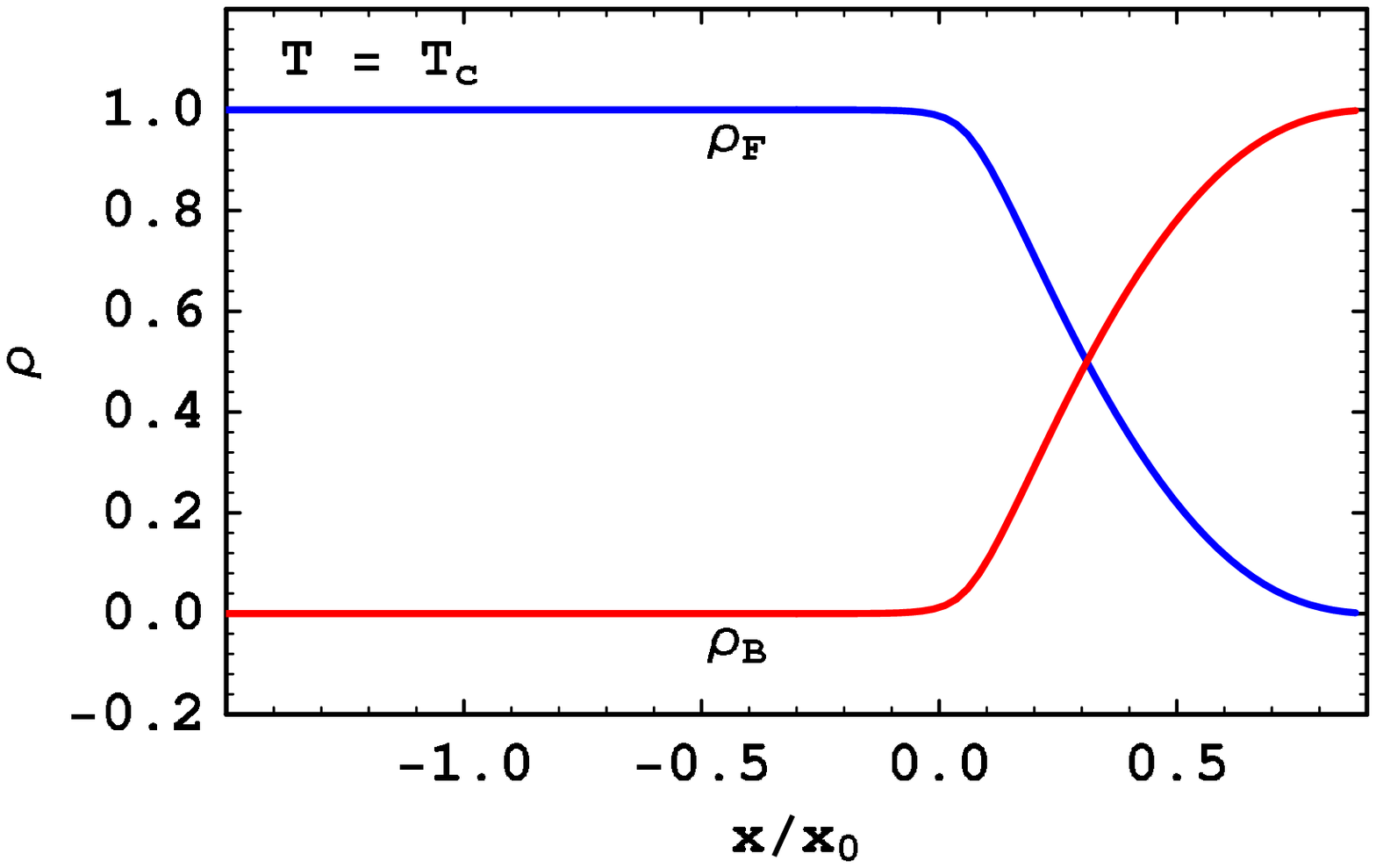}}
%\vspace{0.5cm}
\caption{(Color online) Crossover at the critical temperature.
Left panel: fermion chemical potential $\mu _c$ (blue dotted)
and critical temperature $T_c$ (red dashed) in units of the effective Fermi energy $\e_F$.
Right panel: fermion fraction (blue solid), thermal boson fraction (red solid).}
\label{figTc}
\end{center}
\end{figure*}

The results are shown in the left panel of Fig.\ \ref{figTc}. We see that the chemical potential
behaves qualitatively as for zero temperature. The critical temperature, while
exponentially small in the BCS regime, becomes of the order of and then larger than
the chemical potential during the crossover. This is one of the characteristics of the strong coupling regime
and one reason why this model (in its nonrelativistic version) is used to describe
high-temperature superconductivity \cite{domanski}. In Sec.\ \ref{unitary} we use the ratio
$T_c/\Delta_0$ to illustrate the high-$T_c$ behavior.

The right panel of Fig.\ \ref{figTc} shows the particle density fractions for
fermions and bosons. A crossover similar to the zero-temperature
case can be seen. The density fractions of fermions and bosons suggest that the
crossover is shifted to a slightly larger value of $x$ compared to the zero-temperature case.
While at zero temperature $n_F = n_0$ occurs at $x/x_0 \simeq 0$, here we have
$n_F=n_B$ at $x/x_0 \simeq 0.3$.
It is clear that there is no Bose condensate at $T=T_c$;
the bosonic population rather consists of thermal molecules.
These are uncondensed, strongly-coupled Cooper pairs (see Ref.\ \cite{Chen:2005} for a recent discussion
of this effect in the context of cold atoms). We see that uncondensed pairs do not exist
in the BCS limit. In this case, the superfluid phase transition occurs
``abruptly'', with pair formation and condensation at the same temperature.

In the right panel of Fig.\ \ref{figenergy} we show the fermion and boson energy densities. These curves
are obtained by inserting the solutions for $\mu_c$ and $T_c$ into Eq.\ (\ref{energyfermi}) and
(\ref{energybose}) and making use of $\Delta=0$. We see that, in contrast to
the zero-temperature case, the fermionic energy density increases with $x$ despite $n_F\to 0$. This
is easy to understand from the corresponding quasi-fermion excitations. For all temperatures, the vanishing
chemical potential renders quasi-particles and quasi-antiparticles degenerate. Whereas at $T=0$ they
are both gapped by $\Delta$, at $T=T_c$ they are gapped only by the fermion mass $m$.
For large values of $x$ we have $T_c\gg m$ and quasi-fermions as well as quasi-antifermions can be thermally
excited.
In this context, it would be interesting to consider the formation of chiral condensates
which may be initiated by the degeneracy of quasi-fermions and quasi-antifermions.
We leave this extension of the model for future studies.

The large increase of the bosonic energy density can be understood in the same way.
Note, however, that, in contrast to the fermion mass, the boson mass decreases with increasing crossover
parameter $x$. This difference, together with the different statistics of bosons and fermions gives
rise to the qualitatively different behavior of $E_{B,r}$ compared to $E_{F,r}$.
The strong increase of antiparticle densities has also been predicted in other models
and has been termed ``relativistic BEC (RBEC)'' \cite{Nishida:2005ds,Abuki:2006dv}.
The relativistic effects have also been studied in Ref. \cite{He:2007kd}.

\subsection{Fixed coupling}
\label{allT}

In the previous two subsections we have presented the solution of Eqs.\ (\ref{density})
and (\ref{gapequation}) along two lines in the $x$-$T$ plane: along the line $T=0$
(Sec.\ \ref{zeroT}) and along the (curved) line $T=T_c$ (Sec.\ \ref{Tcrit}). Now we explore
a third path by fixing the crossover parameter and vary the temperature from zero to values
beyond $T_c$. We shall use $x/x_0=0.2$ which is in the intermediate-coupling regime, where
both fermionic and bosonic populations are present.
For $T<T_c$, we use Eqs.\ (\ref{density}) and (\ref{gapequation}) to determine $\mu$ and $\Delta$.
For $T>T_c$, the gap vanishes, $\Delta=0$, i.e., we have the single equation
(\ref{eq:num-03}) to determine the chemical potential $\mu$.

\begin{figure*} [ht]
\begin{center}
%\hbox{
\includegraphics[width=0.5\textwidth]{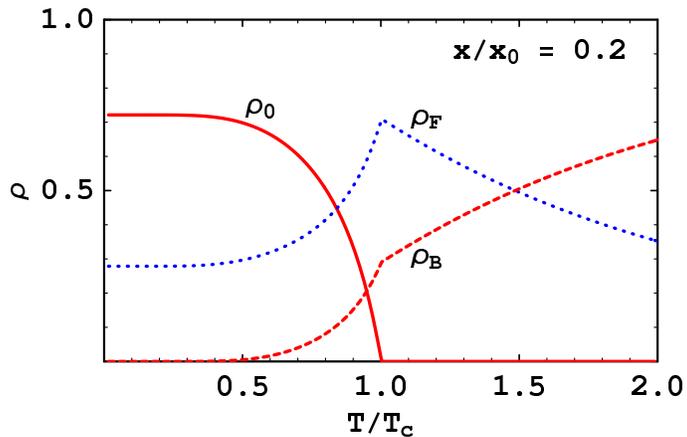}
\caption{(Color online)
Density fractions in the crossover regime at fixed $x/x_0=0.2$
as functions of temperature: condensed bosons (red solid), fermions and uncondensed bosons
(blue dotted and red dashed, respectively).
}
\label{figallT}
\end{center}
\end{figure*}

The condensate and the fermion and boson density fractions are shown in Fig.\ \ref{figallT}.
At the left end, $T=0$, one
recovers the results shown in Fig.\ \ref{figzeroT} at the particular value $x/x_0=0.2$, while
the point $T/T_c=1$ reproduces the respective result shown in Fig.\ \ref{figTc}.
The second-order phase transition manifests itself in
a kink in the density fractions and a vanishing condensate. Below $T_c$ we observe coexistence
of condensed bound states, condensed resonant states, and, for sufficiently large temperatures,
uncondensed bound states. We obtain thermal bosons even above the phase transition.
They can be interpreted as ``preformed'' pairs, just as the uncondensed pairs below $T_c$.
This phenomenon is also called ``pseudogap''
in the literature \cite{Chen:2005,Kitazawa:2005vr}. It suggests that there
is a temperature $T^*(x)$ which marks the onset of pair formation. This
temperature is not necessarily identical to $T_c$. In the BCS regime, $T^*(x)=T_c(x)$,
while for $x\gtrsim 0$, $T^*(x)>T_c(x)$. Of course, our model does not predict any quantitative
value for $T^*$ because thermal bosons are present for all temperatures. Therefore, we expect the model
to be valid only for a limited temperature range above $T_c$.

\subsection{The ratios $\Delta_0/\mu_0$ and $T_c/\Delta_0$}
\label{unitary}

We finally present the results for the ratios $\Delta_0/\mu_0$, and $T_c/\Delta_0$.
They shall serve as a discussion of the dependence of our results on the boson-fermion
coupling $g$ and the fermion mass $m$. Both $g$ and $m$ were fixed throughout the previous sections.
Moreover, we shall see that we reproduce values of these ratios obtained in different models in
certain limit cases.

\begin{figure*} [ht]
\begin{center}
\hbox{\includegraphics[width=0.46\textwidth]{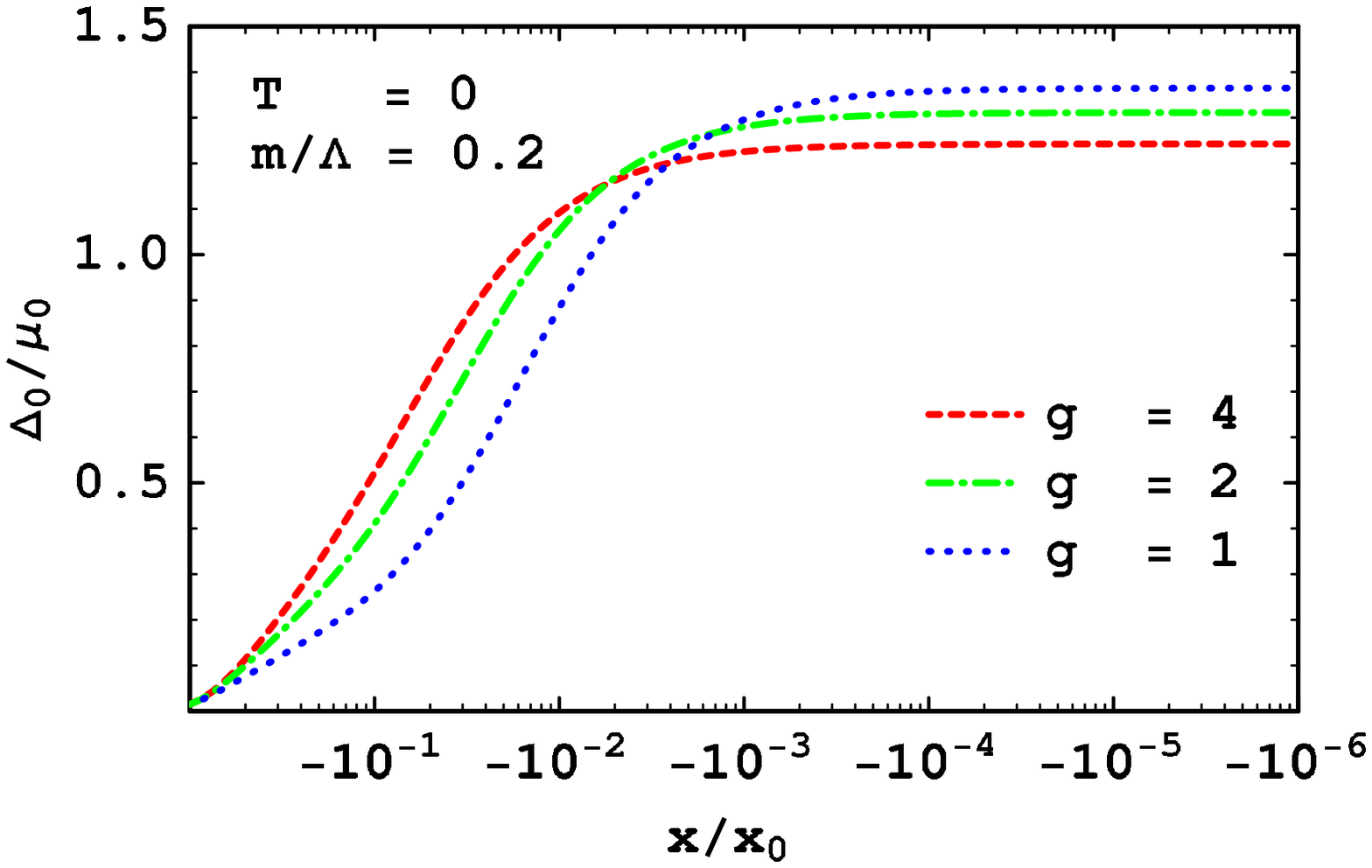}\hspace{0.5cm}
\includegraphics[width=0.46\textwidth]{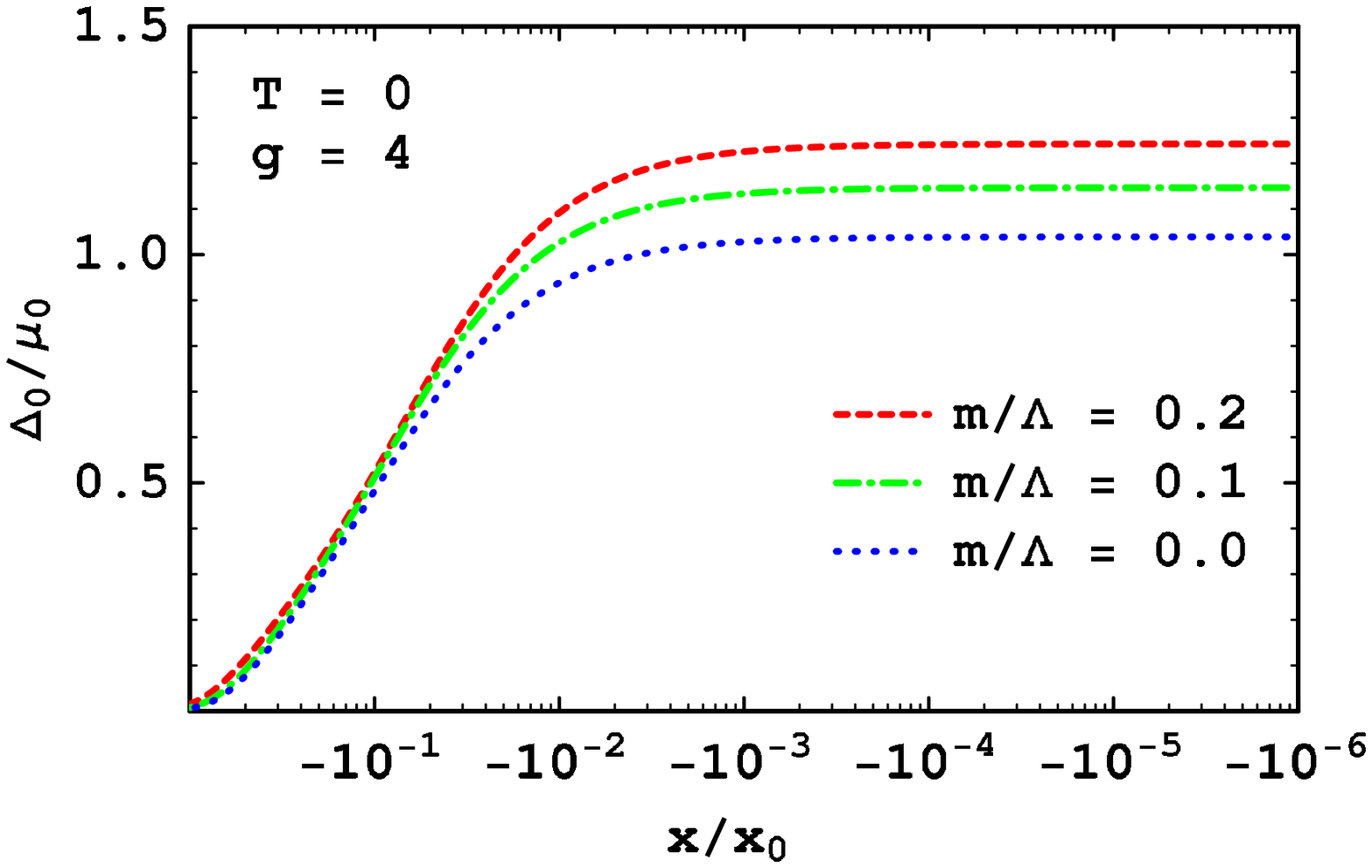}}
%\vspace{0.5cm}
\caption{(Color online) Ratio of gap over chemical potential at zero temperature for
crossover parameters $x/x_0<0$ on a logarithmic scale.
The left end of the horizontal axis corresponds to the BCS regime, the
right end, where $x\to 0$, corresponds to the unitary regime.
Left panel: ratio for different values of the boson-fermion coupling $g$. Right panel:
ratio for different values of the fermion mass.
}
\label{figratio1}
\end{center}
\end{figure*}

Fig.\ \ref{figratio1} shows the ratio $\Delta_0/\mu_0$, using the results for
$\Delta_0$ and $\mu_0$ from Sec.\ \ref{zeroT}. From both panels one can read off the value of the ratio
in the unitary limit, $x\to 0$. For the fermion mass that has been used in the previous subsections,
$m/\Lambda=0.2$, we find $1.2 \lesssim \Delta_0/\mu_0 \lesssim 1.4$. The exact value depends on
the choice of $g$. This range is in agreement with
nonrelativistic, purely fermionic models \cite{chang,Carlson:2005kg,Nishida:2006br}. The right
panel shows that the ratio in the unitary limit decreases with decreasing fermion mass. In particular, in
the ultrarelativistic limit $m=0$ we find $\Delta_0/\mu_0 \simeq 1.0$.

\begin{figure*} [ht]
\begin{center}
\hbox{\includegraphics[width=0.45\textwidth]{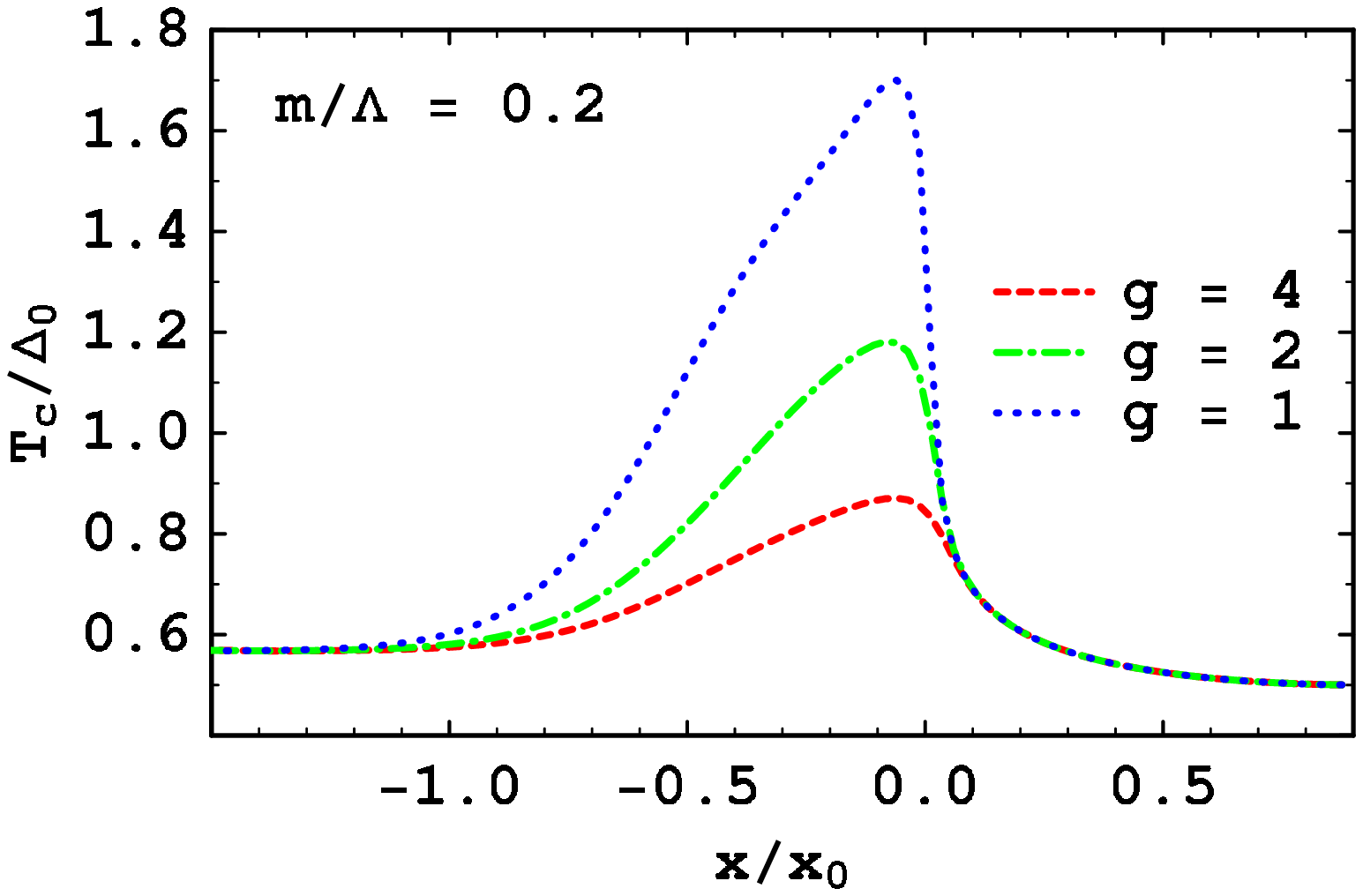}\hspace{0.5cm}
\includegraphics[width=0.45\textwidth]{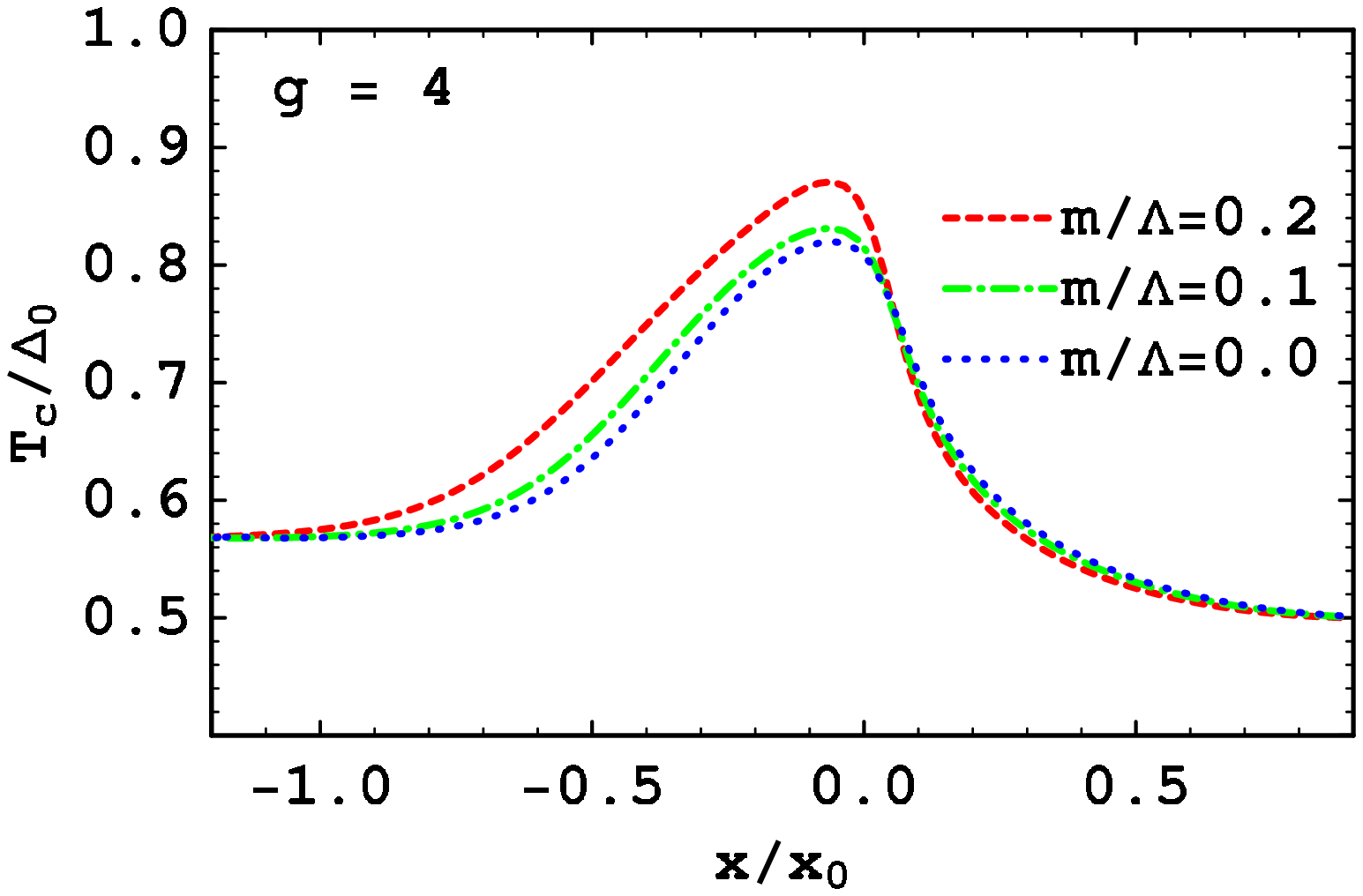}}
%\vspace{0.5cm}
\caption{(Color online) Ratio of critical temperature over zero-temperature gap throughout the
BCS-BEC crossover. Left panel: ratio for different values of the boson-fermion coupling $g$. Right panel:
ratio for different values of the fermion mass.}
\label{figratio2}
\end{center}
\end{figure*}

In Fig.\ \ref{figratio2} we show the ratio $T_c/\Delta_0$, using the result for $T_c$
from Sec.\ \ref{Tcrit}. From BCS theory we know
\be
\lim_{x\to -\infty}\frac{T_c}{\Delta_0} = \frac{e^\g}{\pi} \simeq 0.57 \, ,
\ee
where $\g\simeq 0.577$ is the Euler-Mascheroni constant. This value is reproduced in our results,
independent of $g$ and $m$. Upon increasing the crossover parameter $x$,
the ratio deviates from its BCS value
and increases substantially during the crossover where
it strongly depends on
the coupling $g$. Therefore we make no predictions for its value in the unitary regime. However, we
see that in the BEC regime, the value again becomes independent of the parameters and assumes a value
\be
\lim_{x\to x_0}\frac{T_c}{\Delta_0} \simeq 0.50 \, .
\ee

\section{Two-species system with imbalanced populations}
\label{imbalanced}

It is straightforward to extend our boson-fermion model to two fermion species with
cross-species pairing. This allows
us to introduce a mismatch in fermion numbers and chemical potentials which
imposes a stress on the pairing. This kind of stressed pairing
takes place in a variety of real systems. For example,
quark matter in a compact star is unlikely to exhibit standard BCS pairing in the
color-flavor locked (CFL) phase, i.e., pairing of quarks at a common Fermi surface.
The cross-flavor (and cross-color) pairing pattern of the CFL phase rather suffers a mismatch
in chemical potentials in the pairing sectors $bu-rs$ and $bd-gs$ ($r,g,b$ meaning red, green, blue,
and $u,d,s$ meaning up, down, strange). This mismatch is induced
by the explicit flavor symmetry breaking through the heaviness of the strange quark and by the conditions
of color and electric neutrality. Our system shall only be an idealized and simplified
model of this complicated scenario. However, as in the previous sections, we shall allow for
arbitrary values of the crossover parameter and thus model the strong coupling
regime of quark matter. We shall fix the overall charge and the difference in the
two charges. This is comparable to the effect of
neutrality conditions for matter inside a compact star, which
also impose constraints on the various color and flavor densities.
Our focus will be to find stable homogeneous superfluids in the crossover region and, by
discarding the unstable solutions, identify parameter values where the crossover in fact
becomes a phase transition.
We shall restrict ourselves to zero temperature and defer the full analysis of the
two-species system to a future study \cite{Deng:2006}.

\subsection{Two-fermion system}

We start by replacing the fermion spinor $\psi$ in the Lagrangian (\ref{eq:lag}) with a two-component spinor
\be
\psi = \frac{1}{\sqrt{2}}\left(\begin{array}{c} \psi_1 \\ \psi_2 \end{array}\right) \, ,
\ee
and the chemical potential $\mu$ with the matrix ${\rm diag}(\mu_1,\mu_2)$.
The factor $1/\sqrt{2}$ accounts for the same normalization for the total fermion number
as in the case of single fermion species. We assume both
fermions to have the same mass $m$. Cross-species pairing is taken into account in the
interaction part of the Lagrangian ${\cal L}_I$ in (\ref{eq:lagrangian-fluc1})
which is now replaced by
%\begin{subequations}
\bea
%\mathcal{L}_{f} &=& \frac 12 \overline{\psi}\,[i\g^\mu\partial_\mu+\g_0\,{\rm diag}(\mu_1,\mu_2)-m]\psi
%\, , \\
\mathcal{L}_{I} &=& g\,(\varphi\overline{\psi}_{C}i\gamma_{5}\sigma_1\psi+\varphi^*
\overline{\psi}i\gamma_{5}\sigma_1\psi_{C}) \, ,
\eea
%\end{subequations}
where the Pauli matrix $\sigma_1$ is a matrix in the two-species space.
We denote the average chemical potential and the mismatch in chemical potentials by
\be
\overline{\mu}\equiv\frac{\mu_{1}+\mu_{2}}{2} \, , \qquad
\delta\mu\equiv\frac{\mu_{1}-\mu_{2}}{2}
 \, .
\ee
Then, the bosonic chemical potential is
\be
\mu_{b}=2\overline{\mu} \, .
\ee
The thermodynamic potential differs from the one-fermion case in the dispersion relation
for the fermions,
\begin{eqnarray}
\Omega & = & \frac{m_{b}^{2}-\mu_{b}^{2}}{4g^{2}}\Delta^{2}+\frac{1}{2}\sum_{e}
\int\frac{d^{3}k}{(2\pi)^{3}}\left[\omega_{k}^{e}+2T\ln\left(1-e^{-\omega_{k}^{e}/T}\right)\right]\nonumber \\
 &  & -\sum_{e}\int\frac{d^{3}k}{(2\pi)^{3}}
\left\{ \epsilon_{k}^{e}+T\ln\left[1+e^{-(\epsilon_{k}^{e}+\delta\mu)/T}\right]
+T\ln\left[1+e^{-(\epsilon_{k}^{e}-\delta\mu)/T}\right]\right\} \, ,\label{eq:potential-mis}
\end{eqnarray}
where
\be
\epsilon_{k}^{e}=\sqrt{(\epsilon_{k0}-e\overline{\mu})^{2}+\Delta^{2}}
\ee
and $\omega_{k}^{e}$ as in Eq.\ (\ref{omegak}).
At zero temperature $\Omega$ becomes
\begin{eqnarray}
\Omega & = & \frac{m_{b}^{2}-\mu_{b}^{2}}{4g^{2}}\Delta^{2}-\sum_{e}\int\frac{d^{3}k}{(2\pi)^{3}}\left[\epsilon_{k}^{e}+(\delta\mu-\epsilon_{k}^{e})\Theta(\delta\mu-\epsilon_{k}^{e})\right] \, .
\end{eqnarray}
The particle number densities for each species are derived from the thermodynamic
potential,
\begin{eqnarray}
n_{i} & = & -\frac{\partial\Omega}{\partial\mu_{i}}=\frac{n_{0}}{2}+n_{f,i},\qquad (i=1,2)\, ,
\end{eqnarray}
where $n_{0}$ is given by Eq. (\ref{condensate}) and
\begin{eqnarray}
n_{f,1/2} & = & \sum_{e}e\int\frac{d^{3}k}{(2\pi)^{3}}\left[\frac{\epsilon_{k}^{e}-\xi_{k}^{e}}{2\epsilon_{k}^{e}}\pm e\Theta(\delta\mu-\e_{k}^{e})\frac{\epsilon_{k}^{e}\pm e\xi_{k}^{e}}{2\epsilon_{k}^{e}}\right]\, .
\label{eq:nf12}
\end{eqnarray}
We shall evaluate the model for fixed sum and difference of the particle number densities
\begin{subequations}\label{eq:number-mis12}
\begin{eqnarray}
\overline{n} & \equiv & n_{1}+n_{2}=-\frac{\partial\Omega}{\partial\overline{\mu}}=n_{0}-\sum_{e}\int\frac{d^{3}k}{(2\pi)^{3}}\frac{e\xi_{k}^{e}}{\epsilon_{k}^{e}}\Theta(\epsilon_{k}^{e}-\delta\mu),\label{eq:number-mis1}\\
\delta n & \equiv & n_{1}-n_{2}=-\frac{\partial\Omega}{\partial\delta\mu}=\sum_{e}\int\frac{d^{3}k}{(2\pi)^{3}}\Theta(\delta\mu-\epsilon_{k}^{e})\, .
\label{eq:number-mis2}
\end{eqnarray}
\end{subequations}
Of course, the bosons contribute equally to both particle numbers and thus do not appear in $\delta n$.
The gap equation becomes
\begin{eqnarray}
-x & = & \sum_{e}\int\frac{d^{3}k}{(2\pi)^{3}}
\left[\Theta(\epsilon_{k}^{e}-\delta\mu)\frac{1}{2\epsilon_{k}^{e}}-\frac{1}{2\epsilon_{k0}}\right] \, .
\label{eq:gapeq1}
\end{eqnarray}
We shall solve Eqs.\ (\ref{eq:number-mis12}) and (\ref{eq:gapeq1}) for the variables
$\bar{\mu}$, $\delta\mu$, and $\Delta$.

\subsection{Possible Fermi surface topologies}

\begin{figure}
\includegraphics[scale=0.48]{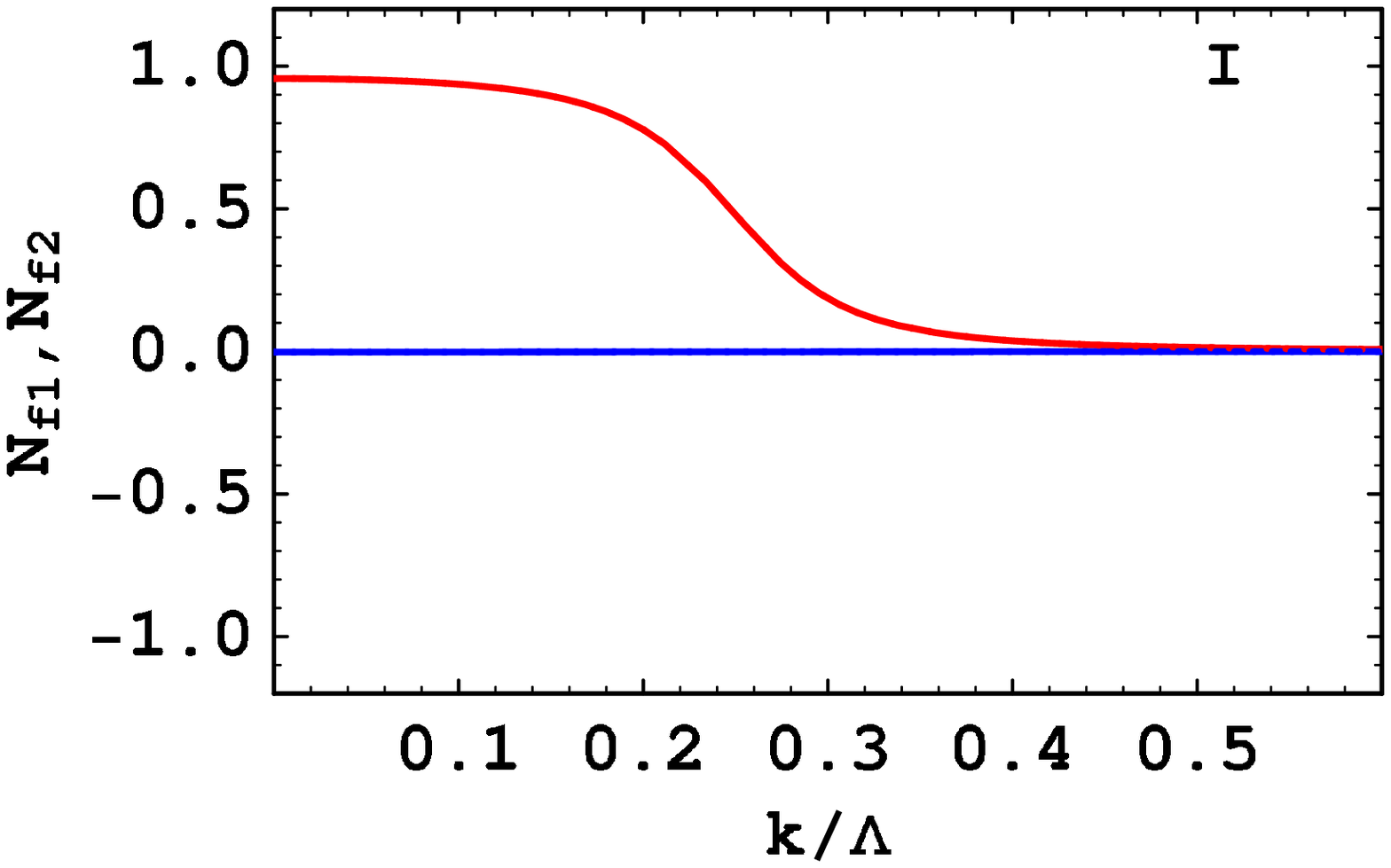}\hspace{0.5cm} 
\includegraphics[scale=0.48]{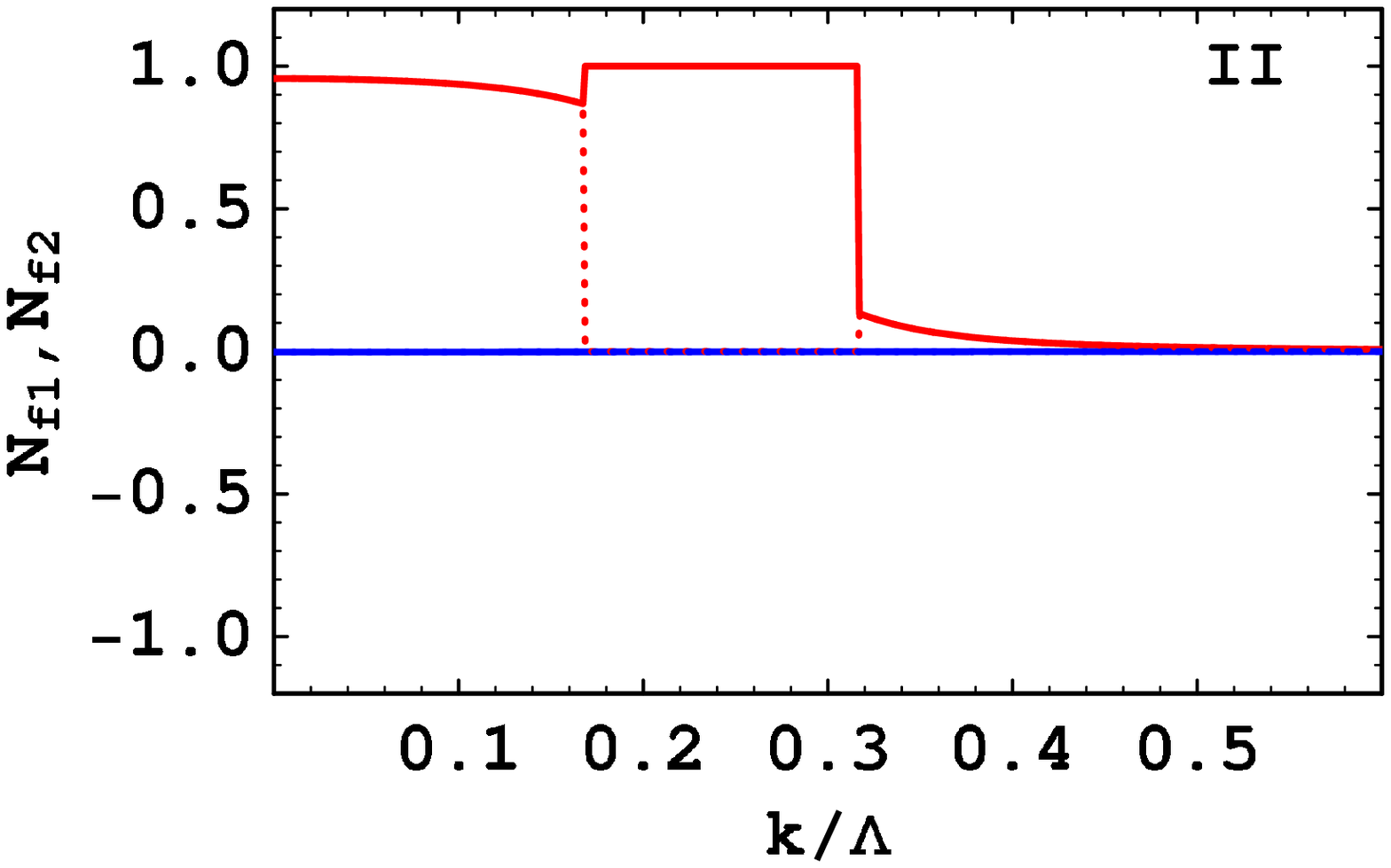} \vspace{0.3cm}\\
\includegraphics[scale=0.48]{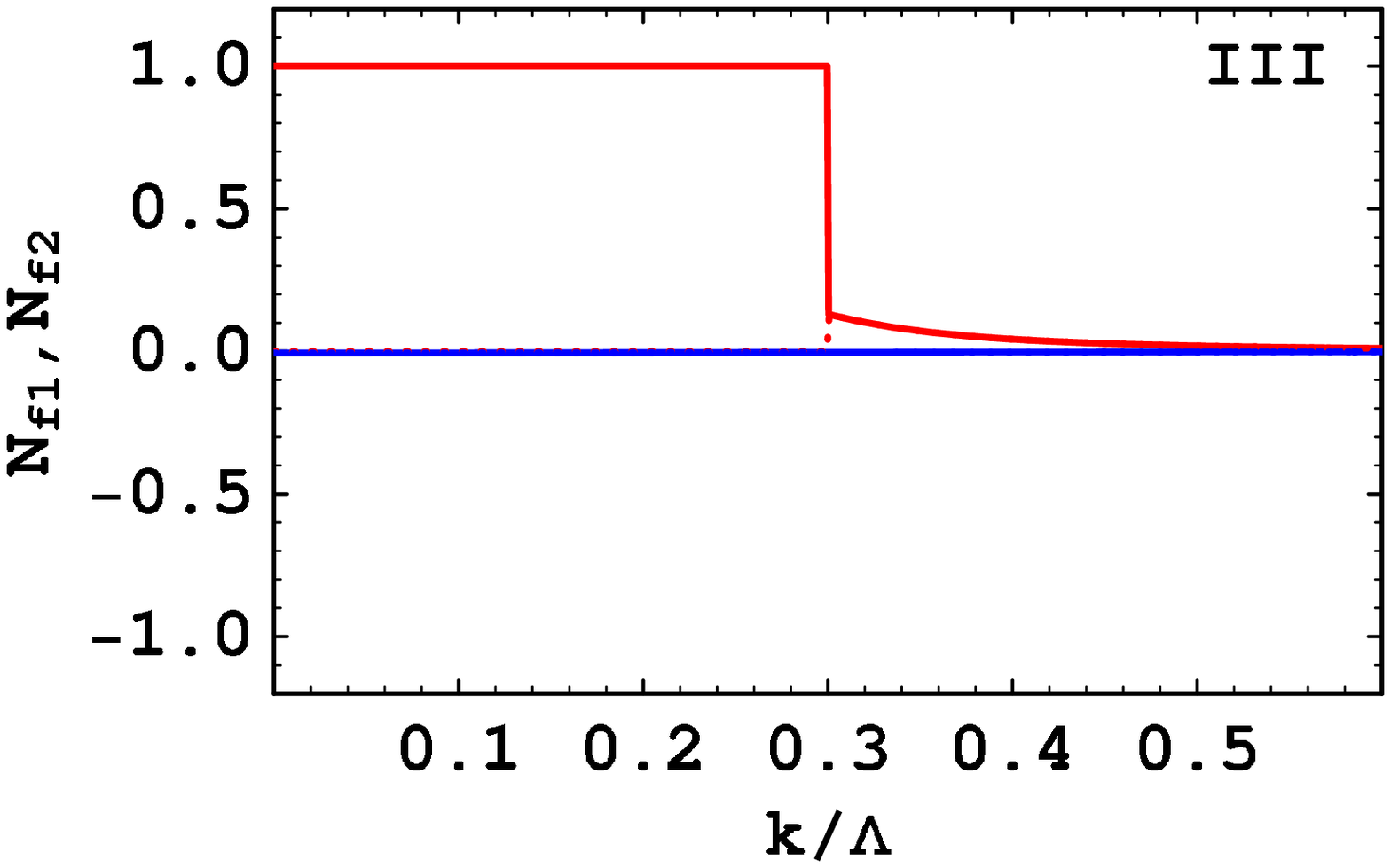}\hspace{0.5cm}
\includegraphics[scale=0.48]{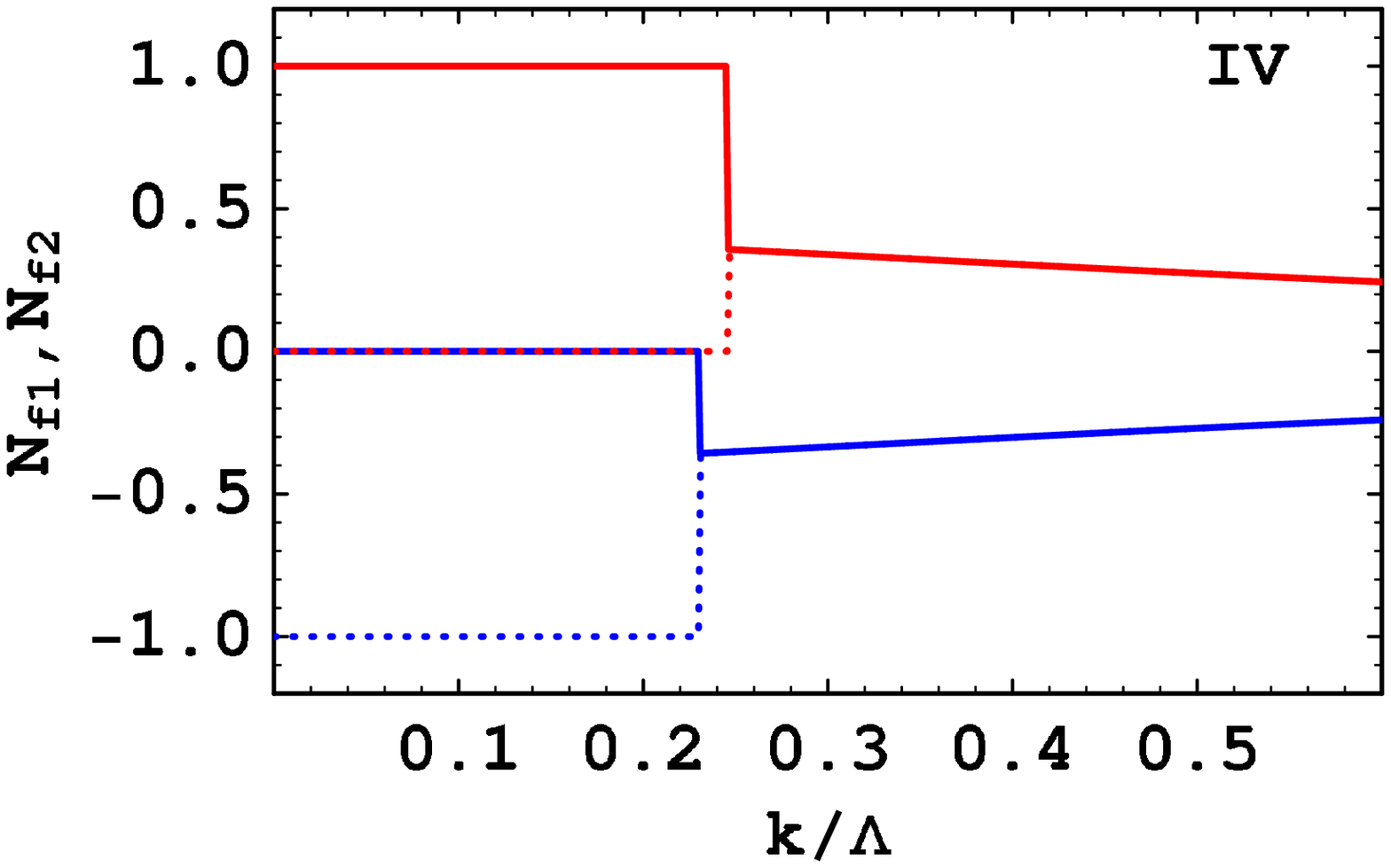}
\caption{\label{cap:occ-number} (Color online) Occupation numbers for fermions (red) and
anti-fermions (blue). Solid and dotted
lines are for species 1 and 2 respectively. The panels illustrate the qualitatively different
cases I through IV from Table \ref{cap:conf}. Note that dotted and solid lines coincide in various
regions, e.g., for all $k$ in the upper left panel. See last two columns of Table \ref{cap:conf}
for the precise form of the occupation numbers.}
\end{figure}

Before we solve the equations, let us comment on their structure, in particular the appearance of the
step function. It is convenient to rewrite
the step functions such that their effect can be translated into the boundaries of the
$dk$ integration. Furthermore, it is instructive to discuss the different particle
and antiparticle occupation numbers in momentum space with the help of the step functions.
As we shall see, the occupation numbers are discontinuous at the zeros of the
dispersion relation $\e_k^e-\d\m$. We assume without loss of generality that $\overline{\mu},\d\m>0$
(in the numerical solution we ensure this by choosing $\overline{n},\d n>0$).
We abbreviate
\be \label{zetarho}
\zeta_\pm \equiv \overline{\mu}\pm\sqrt{\d\m^2-\Delta^2} \, , \qquad  \rho_\pm\equiv \zeta_\pm^2-m^2
\, .
\ee
Then, the step functions are
\begin{subequations}\label{steps}
\bea
\Theta(\d\m-\e_k^+)&=&\Theta(\d\mu-\Delta)\left[\Theta(\rho_+)\Theta(\sqrt{\rho_+}-k)-\Theta(\zeta_-)
\Theta(\rho_-)\Theta(\sqrt{\rho_-}-k)\right] \, , \\
\Theta(\d\m-\e_k^-)&=&\Theta(\d\mu-\Delta)\Theta(-\zeta_-)\Theta(\rho_-)\Theta(\sqrt{\rho_-}-k) \, .
\eea
\end{subequations}
We see that the step functions only give a contribution if $\d\mu>\Delta$. Three different terms
occur, each corresponding to
a different scenario, distinguished by the topology of the effective Fermi surfaces of
fermions and antifermions. Together with the fully gapped state, these are four possible cases.
We list these cases and their characteristics in Table \ref{cap:conf} and Fig.\ \ref{cap:occ-number}.
The fermion dispersion ($e=+$) has either zero, one, or two zeros. A zero corresponds to an effective
Fermi sphere. In particular, a fully gapped state is characterized by the disappearance of any Fermi surface.
The case of two effective Fermi surfaces is termed ``breached pairing'', following the
usual terminology \cite{Gubankova:2003uj}. The antifermion dispersion ($e=-$), in contrast, can have either
zero or one effective Fermi surfaces. Here, the asymmetry between fermions and antifermions is
given by the choice $\overline{\mu}>0$. Hence there is no breached pairing for antifermions.
We find the interesting possibility of filled
Fermi surfaces for fermions of species 1 and antifermions of species 2, see case IV in
Table \ref{cap:conf} and lower right panel in Fig.\ \ref{cap:occ-number}. In Sec.\
\ref{resultsmismatch} we shall see that this is indeed a stable solution for certain values of
the crossover parameter.

\begin{table}
\begin{tabular}{|c||c|c|c|c|}
\hline
cases & characteristics &
parameter region &
$N_{f,i}^{+}(k)$ (fermions) &
$N_{f,i}^{-}(k)$ (anti-fermions)\tabularnewline
\hline\hline
I&
fully gapped &
$\delta\mu<\Delta$ or $\zeta_{+}<m$&
$N_{f1}^{+}=N_{f2}^{+}=N_{\mathrm{gap}}^{+}$&
$N_{f1}^{-}=N_{f2}^{-}=N_{\mathrm{gap}}^{-}$ \tabularnewline
\hline
II&
\begin{tabular}{c}
breached for fermions\tabularnewline
gapped for anti-fermions\tabularnewline
\end{tabular} &
$m<\zeta_{-}<\zeta_{+}$&
\begin{tabular}{c}
$N_{f1}^{+}=1$, $N_{f2}^{+}=0$, $k\in[\sqrt{\rho_{-}},\sqrt{\rho_{+}}]$\tabularnewline
$N_{f1}^{+}=N_{f2}^{+}=N_{\mathrm{gap}}^{+}$, $k\notin[\sqrt{\rho_{-}},\sqrt{\rho_{+}}]$\tabularnewline
\end{tabular}&
same as case I\tabularnewline
\hline
III&
\begin{tabular}{c}
single EFS for fermions\tabularnewline
gapped for anti-fermions \tabularnewline
\end{tabular} &
$|\zeta_{-}|<m<\zeta_{+}$&
\begin{tabular}{c}
$N_{f1}^{+}=1$, $N_{f2}^{+}=0$, $k\in[0,\sqrt{\rho_{+}}]$\tabularnewline
$N_{f1}^{+}=N_{f2}^{+}=N_{\mathrm{gap}}^{+}$, $k\notin[0,\sqrt{\rho_{+}}]$\tabularnewline
\end{tabular}&
same as case I\tabularnewline
\hline
IV&
\begin{tabular}{c}
single EFS for fermions\tabularnewline
single EFS for anti-fermions \tabularnewline
\end{tabular} &
$m<-\zeta_{-}<\zeta_{+}$&
same as case III&
\begin{tabular}{c}
$N_{f1}^{-}=0$, $N_{f2}^{-}=-1$, $k\in[0,\sqrt{\rho_{-}}]$\tabularnewline
$N_{f1}^{-}=N_{f2}^{-}=N_{\mathrm{gap}}^{-}$, $k\notin[0,\sqrt{\rho_{-}}]$\tabularnewline
\end{tabular}\tabularnewline
\hline
\end{tabular}
\caption{\label{cap:conf}Four possible parameter configurations and corresponding
fermion ($e=+$) and antifermion ($e=-$) occupation numbers $N_{f,i}^{e}(k)$
($i=1,2$). The occupation numbers are defined as the integrand in Eq.\ (\ref{eq:nf12}).
We abbreviate ``effective Fermi surface'' by EFS and
$N_{\mathrm{gap}}^{e}\equiv e(\epsilon_{k}^{e}-\xi_{k}^{e})/(2\epsilon_{k}^{e})$.
The different cases are illustrated in Fig.\ \ref{cap:occ-number}.}
\end{table}

\subsection{Number susceptibilities}

In order to check the gapless states for their stability, we have to compute the number
susceptibility matrix \cite{son,Gubankova:2006gj}
\be
\chi_{ij}\equiv \frac{d n_i}{d\mu_j} \, , \qquad i=1,2 \, .
\ee
 Note that we fix $\overline{n}$ and $\delta n$  (or equivalently $n_1$ and
$n_2$) in our solution. Hence $\chi$ can be regarded as measuring the response of the system to a
small perturbation away from this solution. In particular, a stable solution requires
the mismatch in density to increase for an increasing mismatch in chemical potentials.
Therefore, a negative eigenvalue of this 2$\times$2 matrix indicates the instability of a given solution.
The susceptibility is evaluated upon using
\bea \label{evalchi}
\chi_{ij} &=& \frac{1}{2}\frac{d n_0}{d\mu_j} +
\frac{d n_{f,i}}{d\mu_j}
= \frac{\Delta^2}{2g^2}+\left(\frac{2\bar{\mu}\Delta}{g^2} +
\frac{\partial n_{f,i}}{\partial \Delta}\right) \frac{\partial \Delta}{\partial \mu_j} +
\frac{\partial n_{f,i}}{\partial \mu_j} \, .
\eea
The partial derivatives of the fermion densities with respect to the gap and chemical potentials
are straightforwardly computed with the help of
Eq.\ (\ref{eq:nf12}). The partial derivative of the gap with respect to the chemical potentials
can be computed from the
gap equation. To this end, we take the (total) derivative with respect to $\mu_{1/2}$ on both sides
of Eq.\ (\ref{eq:gapeq1}) and solve the equation for $\partial\Delta/\partial\mu_{1/2}$. The results
for the various terms are given in Appendix \ref{appA}.

\subsection{Stable and unstable gapless superfluids}
\label{resultsmismatch}

\begin{figure}
\includegraphics[scale=0.5]{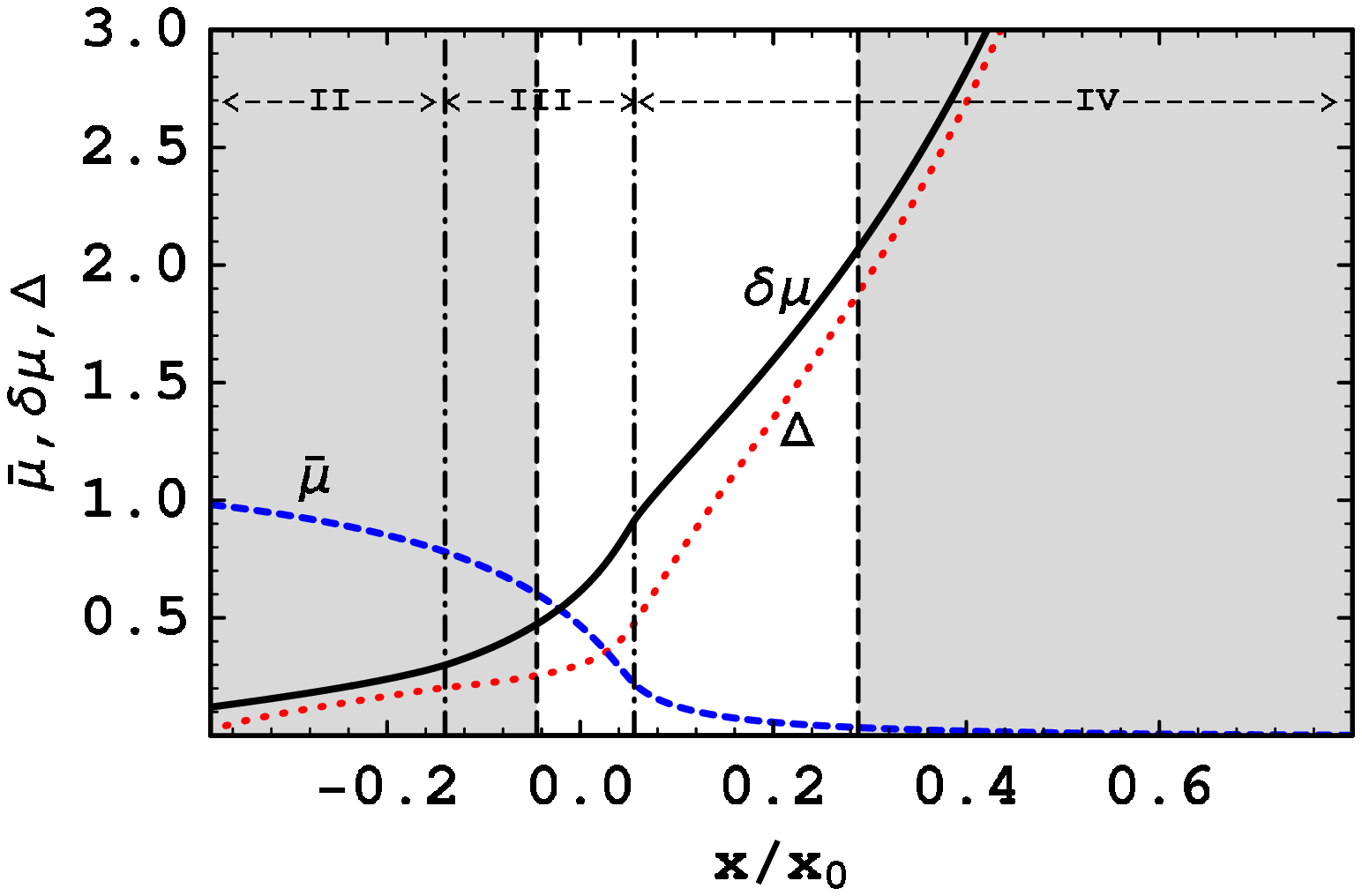}\hspace{0.5cm}
\includegraphics[scale=0.5]{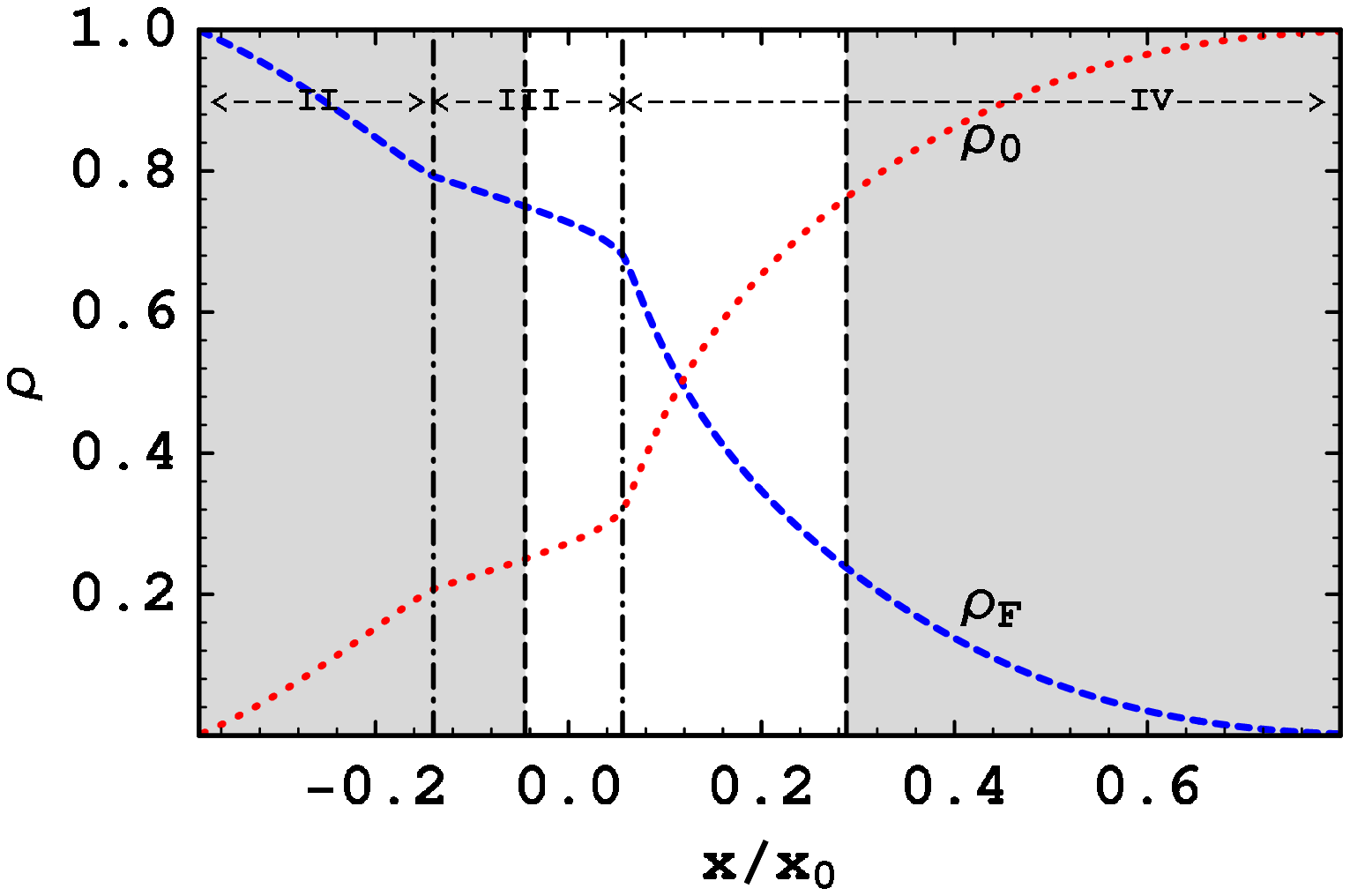}
\caption{\label{cap:mis1} (Color online) The crossover with imbalanced population
at zero temperature. Left panel: $\overline{\mu}$ (blue dashed),
$\delta\mu$ (black solid) and $\Delta$ (red dotted) in units of $\epsilon _F$ for fixed $\overline{n}$
and $\delta n$. Right panel: bosonic (red dotted)
and fermionic (blue dashed) density fractions. The shaded parameter regions are unstable with
respect to a negative susceptibility. The states II, III, and IV correspond to the
respective Fermi surface topologies discussed in Table \ref{cap:conf}. }
\end{figure}

We first present the results for a certain mismatch in densities,
\be
\frac{\delta n}{\overline{n}} = 0.5 \, .
\ee
Moreover, we use $\overline{n}=p_F^3/(3\pi^2)$, and $p_F$, $m$, $g$, as in the main part
of the paper, see Eq.\ (\ref{parameters}). Then, for $T=0$ we have a set of coupled
equations (\ref{eq:number-mis12}) and (\ref{eq:gapeq1}) for any value of the crossover parameter $x$.
We solve this set of equations for $\overline{\mu}$, $\delta\mu$, and $\Delta$.
The results are shown in Fig.\ \ref{cap:mis1}.  The left panel shows that the behavior of the
average chemical
potential and the gap are not unlike the case with a single fermion species, cf.\ Fig.\ \ref{figzeroT}.
Moreover, we see that $\delta\mu >\Delta$ for all $x$. This is clear from Eq.\ (\ref{eq:number-mis2}):
any nonzero $\delta n$ goes along with $\delta\mu >\Delta$. In other words, the standard fully gapped
pairing does not allow (at $T=0$) for a difference in fermion numbers. Thus all solutions correspond
to gapless pairing and case I in Table \ref{cap:conf} and Fig.\ \ref{cap:occ-number} does not appear.
We have indicated in Fig.\ \ref{cap:mis1} for which values of $x$ which of the cases
II, III, and IV occurs. We have also marked the onset of instability. Evaluation of
the susceptibility matrix $\chi(\overline{\mu},\delta\mu,\Delta)$ yields negative eigenvalues
for $x\lesssim -0.04\, x_0\equiv x_-$ and $x\gtrsim 0.28\, x_0\equiv x_+$. In fact,
one of the eigenvalue diverges at these points.
Denoting the two eigenvalues of $\chi$ by $\chi_1$, $\chi_2$, we have
\begin{subequations}
\bea
\chi_1 &\to& \left\{\begin{array}{cc} +\infty & \mbox{for} \quad x \downarrow x_- \, , \; x \uparrow x_+  \\
-\infty & \mbox{for} \quad x \uparrow x_- \, , \; x \downarrow x_+ \end{array}\right. \, ,  \\
\chi_2 &>& 0 \quad \mbox{for all}\; x \, .
\eea
\end{subequations}
Unstable regions of negative $\chi_1$ are shaded in both panels of the
figure. We see that the breached pair solution is always unstable. This is expected from
similar results from mean-field studies for nonrelativistic systems \cite{Gubankova:2006gj}
as well as quark matter (gapless CFL and 2SC phases
\cite{Huang:2004bg}). Interestingly, the stable region consists of two
qualitatively distinct states, labelled by III and IV. The system accounts for a given difference
in number densities not only by filling an effective Fermi
surface with particles of species 1 (state III), but also by additionally filling an anti-Fermi
surface of species 2 (state IV). An interesting
feature of this state is that in the limit of equal Fermi surfaces (1 and anti-2) all charges
($n_1$ and $n_2$) are confined in the Fermi sphere, just as in the unpaired phase. This can
be seen from the lower right panel in Fig.\ \ref{cap:occ-number}: the total occupation numbers
for momenta larger than the effective Fermi momentum vanish since particle and antiparticle contributions
cancel each other. Before this limit is reached, however, this state becomes unstable
for $x> x_+$. This instability in the BEC regime is in contrast to
nonrelativistic systems, where a gapless solution (there with a single effective Fermi surface)
persists throughout the BEC region \cite{son}.

The right panel of Fig.\ \ref{cap:mis1} shows the bosonic and fermionic number density fractions,
cf.\ Fig.\ \ref{figzeroT} for the analogous curves without mismatch. Taking the value of $x$
where $\rho_0 = \rho_F$ as an indicator, we see that the BCS-BEC crossover is shifted to a larger value
of $x$ ($x\simeq 0$ vs.\ $x\simeq 0.1 \, x_0$). Of course, what was a crossover in Fig.\ \ref{figzeroT}
is now actually replaced by at least two phase transitions. The regions marked as unstable will
be replaced by a different phase. It is beyond the scope of this paper to determine these phases, but
it can be expected that they are spatially inhomogeneous, for instance a mixed phase where
a superfluid and normal phases are spatially separated, or some kind of
``Larkin-Ovchinnikov-Fulde-Ferrell'' (LOFF) state \cite{LOFF}. In the stable region we see that
the change in Fermi surface topologies (from state III to IV) is visible in the
boson and fermion fractions which exhibit a kink at this point.

\begin{figure}
\includegraphics[scale=0.5]{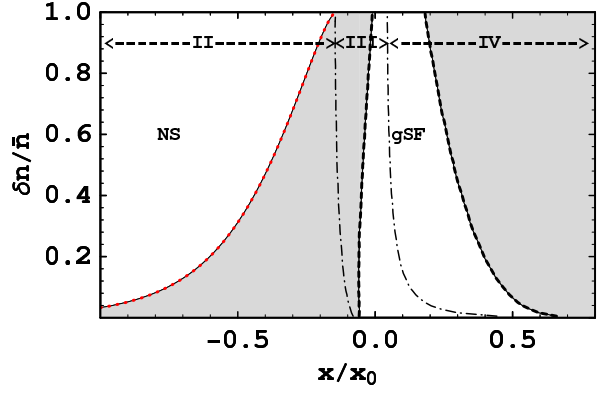}
\caption{\label{cap:phase} The phase diagram in the plane of
the crossover parameter $x$ and the density difference $\delta n/\overline{n}$.
Shaded areas have unstable homogeneous solutions with negative number susceptibility.
NS denotes ``normal state'';
in this region, no solution for the gap equation is found. gSF denotes ``gapless superfluid''; in this
region a stable gapless superfluid state is found with two different Fermi surface topologies, divided
by the right dashed-dotted line. The labels II, III, and IV refer to the states listed in Table
\ref{cap:conf}.}
\end{figure}

We finally present a phase diagram in Fig.\ \ref{cap:phase} for arbitrary (positive) values of
$\delta n/\overline{n}$. Since we do not consider spatially inhomogeneous phases, this phase diagram
is incomplete.
Its main point is to identify regions where homogeneous gapless superfluids may exist.
We find that for sufficiently large mismatches, $\delta n/\overline{n}\gtrsim 0.02$ there is a region
where no solution with nonzero $\Delta$ can be found (we have not shown this region in the above
results for $\delta n/\overline{n} = 0.5$). We see that the region of stable superfluids shrinks with
increasing
$\delta n/\overline{n}$, as expected. Note that the horizontal axis $\delta n/\overline{n}=0$
is not continuously connected
to the rest of the phase diagram. For vanishing mismatch in densities a stable, fully gapped
superfluid exists for all $x$, as we saw in the main part of the paper. One should thus not be
misled by the instability for arbitrarily small mismatches.

We conclude with emphasizing the two main qualitative differences to
analogous phase diagrams in nonrelativistic systems: $(i)$ within the stable region of homogeneous
gapless superfluids there is a curve that separates two different Fermi surface topologies;
this is the right dashed-dotted line in Fig.\ \ref{cap:phase}. $(ii)$ for large $x$ the gapless superfluid
becomes unstable even in the far BEC region; this is the shaded area on the right side in
Fig.\ \ref{cap:phase}.

%\begin{figure}
%\includegraphics[scale=0.5]{susceptibility1-x.eps}
%\includegraphics[scale=0.5]{susceptibility2-x.eps}
%\caption{\label{cap:mis2} Eigenvalues for the number susceptibility matrix
%$\chi$ for the solution shown in Fig. \ref{cap:mis1}.}
%\end{figure}

\section{Summary and outlook}

We have studied the relativistic BCS-BEC crossover for zero and nonzero temperatures
within a boson-fermion model. Variations of this model have previously been used for
nonrelativistic systems in order
to study cold fermionic atoms and high-temperature superconductors. The bosons of the model
are bound states of fermion pairs. Conversion of two fermions into a boson and vice versa
is implemented by requiring chemical equilibrium with respect to this process.
The crossover is realized by varying an effective coupling strength $x$, constructed
from the difference between the renormalized boson mass $m_{b,r}$ and the boson chemical potential
$\mu_b$, and the boson-fermion coupling constant $g$, $x=-(m_{b,r}^2-\mu_b^2)/(4g^2)$.
In this form, $1/x$ plays the role of the scattering length,
in particular $1/x=\pm \infty$ in the unitary limit. We have evaluated the model in its simplest
form, employing a mean-field approximation.

An important property of the model is the coexistence of weakly-coupled
Cooper pairs with condensed and uncondensed bosonic bound states. In the crossover regime
as well as in the BEC regime, strongly-bound molecular Cooper pairs exist below and above
the critical temperature $T_c$. Above $T_c$, they are all uncondensed (``preformed'' Cooper pairs)
while below $T_c$ a certain fraction of them forms a Bose-Einstein condensate. In contrast, in the BCS
regime, pairing and condensation of fermionic degrees of freedom (in the absence of bosons) both
set in at $T_c$.

Furthermore, we have characterized the onset of nonzero antifermion and antiboson populations during the
crossover by computing the energy density. The reason for the appearance of antiparticles
is the strong decrease of the fermion chemical potential.
While the fermion chemical potential
is identical to the Fermi energy in the BCS regime, it reaches values well below the fermion
mass in the BEC regime. As a consequence, particle and antiparticle excitation energies become
almost identical and thus antiparticles are present for nonvanishing temperatures.

Finally, we have extended the model by considering two fermion species with mismatched
densities. This case has been evaluated for zero temperature. We have found
stable gapless superfluids in the crossover region. In contrast to nonrelativistic systems,
we found no stable homogeneous phase in the far BEC region. Moreover, two different stable
Fermi surface configurations have been identified. Besides a state with a single effective Fermi surface,
also found in nonrelativistic systems, we found the possibility of a superfluid phase with
Fermi surfaces for particles of species 1 and anti-particles of species 2. A complete evaluation
of the two-species model, including inhomogeneous phases and nonzero temperatures,
remains to be done in the future.

The model may be extended in several ways, in order to describe more realistic scenarios,
for instance dense quark matter in the interior of a compact star. First, one may go beyond
the mean field approximation, which seems particularly interesting in the crossover region, where
the validity of this approximation is questionable. Also, one may introduce more than two
fermion species, accounting for color and flavor degrees of freedom in quark matter.
Finally, we propose to include chiral condensates into the model.

\begin{acknowledgments} A.S.\ acknowledges valuable discussions with M.\ Alford, S.\ Reddy,
I.\ Shovkovy, and
support by the U.S. Department of Energy under contracts DE-FG02-91ER50628
and DE-FG01-04ER0225 (OJI). Q.W.\ thanks P.-f. Zhuang for helpful discussions
and is supported in
part by the startup grant from University of Science and Technology of China (USTC) in
association with 'Bai Ren' project of Chinese Academy of Sciences (CAS) and by
National Natural Science Foundation of China (NSFC) under the grant 10675109.

\end{acknowledgments}

\appendix

\section{Calculation of number susceptibilities}
\label{appA}

In this appendix, we compute the elements of the number susceptibility matrix as given in
Eq.\ (\ref{evalchi}).
For a compact notation we introduce the following abbreviations for integrals containing
a $\d$-function
\begin{subequations}
\bea
\d_1^\pm&\equiv&
\sum_e\int\frac{d^3k}{(2\pi)^3}\frac{\Delta}{\e_k^e}\frac{\e_k^e\pm e\xi_k^e}{2\e_k^e}\d(\d\m-\e_k^e) \, ,
\\
\d_2^\pm&\equiv&
\sum_e \int\frac{d^3k}{(2\pi)^3}\left(\frac{\e_k^e\pm e\xi_k^e}{2\e_k^e}\right)^2\d(\d\m-\e_k^e) \, ,
\\
\d_3&\equiv& \sum_e \int\frac{d^3k}{(2\pi)^3}\frac{\Delta^2}{4(\e_k^e)^2}\d(\d\m-\e_k^e) \, ,
\eea
\end{subequations}
and integrals containing a step function
\begin{subequations}
\bea
\Theta_1&\equiv&
\sum_e\int\frac{d^3k}{(2\pi)^3}\frac{e\Delta\xi_k^e}{2(\e_k^e)^3}\Theta(\e_k^e-\d\m) \, ,
\\
\Theta_2&\equiv&
\sum_e \int\frac{d^3k}{(2\pi)^3}\frac{\Delta^2}{4(\e_k^e)^3}\Theta(\e_k^e-\d\m) \, .
\eea
\end{subequations}
In this notation the various terms in Eq.\ (\ref{evalchi}) become
\be
\frac{\partial n_{f,1/2}}{\partial\Delta} = \Theta_1\mp\d_1^\pm \, ,
\ee
and
\be
\frac{\partial n_{f,1/2}}{\partial\mu_{1/2}}= \Theta_2+\d_2^\pm\, , \qquad
\frac{\partial n_{f,1/2}}{\partial\mu_{2/1}}= \Theta_2-\d_3 \, .
\ee
The partial derivative of the gap with respect to the chemical potentials, obtained from the
gap equation, is
\bea
\frac{\partial\Delta}{\partial\mu_{1/2}}&=& \frac{2\bar{\mu}\Delta/g^2+\Theta_1\mp\d_1^\pm}{4(\Theta_2-\d_3)}
 \, .
\eea
Consequently,
\begin{subequations} \label{chiij}
\bea
\chi_{11/22} &=& \frac{\Delta^2}{2g^2}+\Theta_2+\delta_2^\pm +
\frac{(2\bar{\mu}\Delta/g^2+\Theta_1\mp\d_1^\pm)^2}{4(\Theta_2-\d_3)} \\
\chi_{12} &=& \chi_{21} = \frac{\Delta^2}{2g^2}+\Theta_2-\delta_3 +
\frac{(2\bar{\mu}\Delta/g^2+\Theta_1-\d_1^+)(2\bar{\mu}\Delta/g^2+\Theta_1+\d_1^-)}{4(\Theta_2-\delta_3)}
\eea
\end{subequations}
In particular, we see that the susceptibility matrix is symmetric.

We may further evaluate the terms with the $\delta$-functions.
For any function $f^e(k)$ we have
\bea
\sum_e\int\frac{d^3k}{(2\pi)^3} f^e(k)\delta(\d\m-\e_k^e)
&=&\frac{\Theta(\d\m-\Delta)}{2\pi^2}\frac{\d\m}{\sqrt{\d\m^2-\Delta^2}}\left\{\Theta(\rho_+)
\sqrt{\rho_+}\,\zeta_+ \,f^+(\sqrt{\rho_+})\right.\non
&& \left. +\,\Theta(\rho_-)
\sqrt{\rho_-}\,\zeta_-[\Theta(\zeta_-)f^+(\sqrt{\rho_-})
-\Theta(-\zeta_-)f^-(\sqrt{\rho_-})]\right\} \, ,
\eea
with $\zeta_\pm$ and $\rho_\pm$ defined in Eq.\ (\ref{zetarho}).
Consequently,
\begin{subequations} \label{deltas}
\bea
\d_1^\pm
&=& \frac{1}{2\pi^2}\frac{\Delta\Theta(\d\m-\Delta)}{2\d\m\sqrt{\d\m^2-\Delta^2}}
\left[\Theta(\rho_+)\sqrt{\rho_+}\,\zeta_+ \,h_\pm+ \Theta(\rho_-)\sqrt{\rho_-}
\,\zeta_-\,h_\mp\,{\rm sgn}\zeta_-\right] \, ,
\\
\d_2^\pm
&=& \frac{1}{2\pi^2}\frac{\Theta(\d\m-\Delta)}{4\d\m\sqrt{\d\m^2-\Delta^2}}\left[\Theta(\rho_+)\sqrt{\rho_+}
\,\zeta_+\,h_\pm^2 + \Theta(\rho_-)\sqrt{\rho_-}\,\zeta_-\,h_\mp^2\,{\rm sgn}\zeta_-\right] \, ,
\\
\d_3
&=& \frac{1}{2\pi^2}\frac{\Delta^2\Theta(\d\m-\Delta)}{4\d\m\sqrt{\d\m^2-\Delta^2}}
\left[\Theta(\rho_+)\sqrt{\rho_+}\,\zeta_+
+ \Theta(\rho_-)\sqrt{\rho_-}\,\zeta_-\,{\rm sgn}\zeta_-\right] \, ,
\eea
\end{subequations}
where we abbreviated
\be
h_\pm\equiv \d\m\pm\sqrt{\d\m^2-\Delta^2} \, .
\ee
For the integrals with the step function we make use of Eqs.\ (\ref{steps}) to find for any
function $f^e(k)$
\bea \label{anystep}
\sum_e\int\frac{d^3k}{(2\pi)^3} f^e(k)\Theta(\e_k^e-\d\m) &=& \sum_e\int\frac{d^3k}{(2\pi)^3} f^e(k)
-\frac{\Theta(\d\m-\Delta)}{2\pi^2}\left[\Theta(\rho_+)\int_0^{\sqrt{\rho_+}}dk\,k^2f^+(k)\right.\non
&&\left.-\;\Theta(\zeta_-)\Theta(\rho_-)\int_0^{\sqrt{\rho_-}}dk\,k^2f^+(k)+
\Theta(-\zeta_-)\Theta(\rho_-)\int_0^{\sqrt{\rho_-}}dk\,k^2f^-(k)\right] \, .
\eea
We insert Eqs.\ (\ref{deltas}) and Eq.\ (\ref{anystep}) (the latter with the respective integrand replacing
$f^e(k)$) into Eq.\ (\ref{chiij}) to evaluate the susceptibility matrix $\chi$.

\end{document}